\def\isarxiv{1}
\ifdefined\isarxiv
\documentclass[smallextended]{svjour3}
\else
\documentclass[smallextended]{svjour3}
\fi
\usepackage[british]{babel}
\addto\captionsbritish{\renewcommand{\figurename}{Fig.}}
\usepackage{microtype}
\usepackage[round,sort&compress]{natbib}
\let\cite\citep
\usepackage{graphicx}
\usepackage{xcolor}
\ifdefined\highlightrevisions
	\definecolor{revisioncolour}{RGB}{0,76,153}
\else
	\definecolor{revisioncolour}{RGB}{0,0,0}
\fi
\newcommand{\changed}[1]{{\color{revisioncolour}#1}}
\usepackage{url}
\usepackage{array}
\usepackage{booktabs}
\usepackage{framed}

\makeatletter
\let\cl@chapter\undefined
\makeatother
\usepackage{cleveref}

\journalname{Empirical Software Engineering}
\renewcommand{\makeheadbox}{}
\date{Accepted for publication in Empirical Software Engineering}

\begin{document}

\ifdefined\isarxiv
\title{How the Misuse of a Dataset Harmed Semantic Clone Detection\thanks{\changed{For the purpose of open access, the authors have applied a Creative Commons (CC BY) licence to any Author Accepted Manuscript version arising.}}}
\else
\title{How the Misuse of a Dataset Harmed Semantic Clone Detection\thanks{For the purpose of open access, the authors have applied a Creative Commons (CC BY) licence to any Author Accepted Manuscript version arising.}}
\fi

\author{Jens Krinke, Chaiyong Ragkhitwetsagul}

\institute{J.~Krinke \at
  UCL Computer Science, University College London, London, UK\\
  \email{j.krinke@ucl.ac.uk} 
\and
  C.~Ragkhitwetsagul \at
  Faculty of Information and Communication Technology (ICT), 
  Mahidol University, Nakhon Pathom, Thailand \\
  \email{chaiyong.rag@mahidol.ac.th}
}

\maketitle

\begin{abstract}
BigCloneBench is a well-known, widely used large-scale dataset for evaluating the recall of traditional clone detection tools.
It has been beneficial for research on clone detection and has become a standard in evaluating the performance of clone detection tools.
More recently, it has also been widely adopted as a dataset to evaluate machine learning approaches to semantic clone detection or to detect functionally similar code.

This paper demonstrates that the Weakly Type-3/Type-4 (WT3/T4) subset of BigCloneBench is problematic as ground truth for learning or evaluating functional similarity, and highlights aspects of its construction that severely compromise its quality for this specific task.
\changed{BigCloneBench's \texttt{tagged-tagged} WT3/T4 clone pairs combine two methods labelled as true positives for the same functionality and do not contain exemplar functions.}
\changed{A population-weighted analysis of a stratified probability sample of 406 such pairs estimated that 93.9\% of pairs in the eligible \texttt{tagged-tagged} WT3/T4 population are not functionally similar under our operational definition and are therefore mislabelled (conservative 95\% confidence interval: 91.5\%--96.2\%).}

From a literature review of 179 papers that use BigCloneBench as a dataset, we estimate that 128 papers used BigCloneBench to evaluate semantic clone detection in a way that their results could be affected by the mislabelling.
\changed{These papers often report high F1 scores (e.g., above 0.9), but such scores on BigCloneBench do not establish whether an approach has learned functional similarity or relied on dataset-specific shortcuts.}

\changed{Our study does not assess BigCloneBench's validity for evaluating syntactic or textual clone detection of Type-1, Type-2, and Type-3 clones.}
\changed{We acknowledge BigCloneBench's important contributions to more than a decade of traditional clone detection research.}
\changed{However, using BigCloneBench's WT3/T4 subset as ground truth for same-main-function similarity without validating its labels for that task constitutes misuse. By threatening the validity of a substantial body of reported results, this practice has harmed the evidence base of semantic clone detection.}
We encourage a more critical and responsible use of benchmarks and datasets and ask for more rigorous validation during dataset creation and use in the future.
\end{abstract}


\section{Introduction}

The performance evaluation of clone detection or code similarity detection approaches is a common problem that needs well-constructed benchmarks.
There have been a series of benchmarks that have been used in the field of clone detection, e.g., by \citet{Bellon2007}, \citet{Krutz2014}, \citet{Ragkhitwetsagul2018}, or \citet{Kamp2019}.
However, many benchmarks are of a limited scale due to the effort required in manually constructing the benchmarks.
BigCloneBench~\citep{Svajlenko2014,Svajlenko2015,Svajlenko2017,Svajlenko2021} is an exception, as it is a large-scale dataset of method-level clones mainly targeted at the evaluation of the recall of clone detection tools.
Many papers have used BigCloneBench to evaluate and compare the performance of clone detection tools, and it has become standard in the field of traditional clone detection to evaluate the recall with BigCloneBench, together with a manual evaluation of a sample to evaluate precision.

Recent approaches to clone and functional similarity detection are based on machine learning and are trained on large-scale datasets.
It is tempting to adopt BigCloneBench as the ground truth for learning functional similarity. 
However, recent work~\citep{Krinke2022} has highlighted that BigCloneBench is problematic for machine learning approaches and for the evaluation of Type-4 Clone Detectors (Type-4 Clones are also known as Semantic Clones).
There are other studies that also investigated BigCloneBench.
\citet{Yu2022} showed that one can achieve high effectiveness for detecting semantic clones when evaluated on BigCloneBench by merely training a linear model to match identifier names.
\citet{Alam2023} note that BigCloneBench contains only a few genuine Java semantic clones and suffers from imbalance and clone-definition ambiguity.
\citet{zhou2025} report a data leakage ratio of 55.7\%, the second worst ratio across 83 software engineering benchmarks.

None of the above papers, including our own, has identified the core problem of BigCloneBench, which is the way in which the ground truth has been constructed.
We will show that BigCloneBench has been created in a semi-automatic way, where human judges only labelled individual methods as containing a specific functionality or not, \emph{even when the functionality is only a minor part of the method}.
The ``true'' clone pairs are then automatically constructed as pairs of methods that share a specific functionality, \emph{even when their overall functionality is very different}.
\changed{This semi-automatic approach creates a flawed WT3/T4 ground truth whose true clone pairs have not been manually validated. Our human study shows that the majority of the sampled pairs are mislabelled under our operational definition.}

Similar problems have been identified in other datasets used for similar purposes.
\changed{An investigation of the CoSQA+ dataset for code search found that 36.7\% of code fragments that satisfy at least one query satisfy two or more, showing how one-to-one ground truths can mislabel valid query-code pairs~\citep{gong2025cosqa}.}
Most recently, OpenAI announced that it would no longer evaluate models on the widely used SWE-bench Verified benchmark due to widespread issues with flawed test cases that reject correct solutions and severe data contamination in the training of frontier models~\citep{openai2026swebench}.
In Computer Vision, the ImageNet dataset was considered the gold standard.
However, recent work has highlighted problems with the dataset, such as label errors, and has shown that these errors can significantly affect the performance of models trained on the dataset~\citep{northcutt2021pervasive,kisel2024flaws,hardt2025emerging}.

\changed{Our research objective is to determine how the construction of BigCloneBench affects its suitability as ground truth for functional similarity.}
\changed{This includes estimating the extent of mislabelling under our operational definition and examining the implications for published research.}
\changed{Our focus is the alignment between the WT3/T4 ground truth and the construct of same-main-function similarity when the dataset is repurposed for semantic clone detection.}
\changed{We address this objective through a sequence in which embeddings provide a scalable diagnostic, a human-labelled stratified probability sample supplies the primary estimate and relates it to the construction process, and an LLM comparison subjects the human consensus labels to post-hoc scrutiny.}
\changed{A literature review then estimates how many papers contain potentially affected results, after which further dataset observations and alternative benchmarks inform the resulting recommendations.}

\changed{We show that, despite its value in traditional clone detection, repurposing BigCloneBench's WT3/T4 subset without validating its labels for same-main-function similarity creates systematic threats to validity.}
\changed{Our literature review estimates that 128 papers contain results whose validity may be affected by this mismatch.}
\changed{We use \emph{misuse} to denote this specific practice: treating WT3/T4 as ground truth for same-main-function similarity without establishing that its labels represent that construct.}
\changed{We use \emph{harmed} to denote the aggregate consequences for the field's evidence base: a substantial body of results has become difficult to interpret and requires re-evaluation before it can provide reliable evidence of progress.}

The contributions of this paper are:
\begin{itemize}
\item \changed{Identification and explanation of the mismatch between BigCloneBench's recall-oriented construction and its use as ground truth for same-main-function semantic similarity.}
\item \changed{An exploratory diagnostic showing that BigCloneBench and TBCCD have lower agreement between ground truth and embedding similarity than six comparison datasets across seven contemporary embedding models.}
\item \changed{A population-weighted estimate from a 406-pair stratified probability sample that 93.9\% of the eligible WT3/T4 sampling population is mislabelled under our operational definition (conservative 95\% confidence interval: 91.5\%--96.2\%).}
\item \changed{An estimate that 128 of 179 reviewed papers contain BigCloneBench-based results whose validity may be affected by the WT3/T4 ground-truth mismatch.}
\item \changed{Eight further observations characterising BigCloneBench's labelling, pair construction, class and functionality distributions, repair attempts, and propagation into the inspected derived datasets.}
\item \changed{A task-oriented catalogue and comparison of alternative semantic-clone benchmarks and datasets, including their principal limitations.}
\end{itemize}

\changed{Accordingly, results that rely on the WT3/T4 subset as ground truth for semantic equivalence may require re-evaluation with a benchmark aligned to that construct.}

\begin{framed}
\noindent\textbf{Problem Statement:}
\changed{BigCloneBench, a dataset originally designed for evaluating \emph{syntactic} clone detection, has been widely repurposed, and its WT3/T4 subset has been misused as ground truth for same-main-function semantic similarity without independent validation of that construct.}
However, the dataset construction does not support reliable ground truth for \emph{functional similarity}. The reason is that pairs of methods are labelled as true clone pairs even when they only share a minor functionality, while their main overall functionality is very different.
\changed{This paper provides empirical evidence of extensive mislabelling among sampled WT3/T4 true clone pairs under our operational definition and demonstrates the implications for the validity of published research.}
\end{framed}

\changed{Our critique concerns the alignment between the dataset's construction and its later use, rather than the intentions of its creators or the conduct of individual researchers.}

The remainder of this paper is structured as follows:
Section~2 provides a brief background on the definitions and classifications of code clones.
Section~3 details the construction process of BigCloneBench, highlighting the semi-automatic pairing mechanism that compromised the semantic ground truth.
\changed{Section~4 presents a diagnostic embedding analysis, a manual investigation of a stratified probability sample, and a post-hoc LLM comparison.}
\changed{Section~5 estimates how many papers contain results potentially affected by WT3/T4 mislabelling through a systematic literature review of papers that used the dataset.}
Section~6 explores further structural problems of the dataset, including extreme class imbalance and the contamination of derived datasets.
\changed{Section~7 discusses the methodological implications and provides recommendations for future use.}
Section~8 reviews alternative benchmarks.
Finally, Section~9 addresses threats to validity, and Section~10 concludes the paper.

\section{A Short Introduction to Clone Detection}

In recent years, the importance of clone detection and code similarity detection has grown significantly.
Clone detection is the process of identifying similar code fragments in a codebase.
The term ``clone'' is used to refer to code fragments that are similar to each other.
However, there is no formal or precise definition of what a clone is, and the usual definition\footnote{This definition stems from the ongoing struggle of the community to come up with a good definition, and Ira D.~Baxter jokingly gave this definition at a workshop.} of a clone pair is \emph{a pair of code fragments that are similar to each other according to some definition of similarity}.

If two code fragments, sometimes called ``snippets'', are similar to each other, they are usually classified into one of four types of clones.
\citet{Svajlenko2014} gives the following definitions for the four types:
\begin{description}
\item[Type-1:] Syntactically identical code snippets, except for differences in white space, layout, and comments.
\item[Type-2:] Syntactically identical code snippets, except for differences in identifier names, literal values, white space, layout, and comments.
\item[Type-3:] Syntactically similar code snippets that differ at the statement level. 
Snippets have statements added, modified, and/or removed with respect to each other.
\item[Type-4:] Syntactically dissimilar code snippets that implement the same functionality.
\end{description}
Note that types 1–3 are usually referred to as syntactic clones, as they are textual or syntactic variants of each other.
Type-4 clones are usually referred to as semantic clones, as they are functionally similar to each other, even if they are syntactically different.
The process of identifying semantic clones is called semantic clone detection.

The definitions of the four types vary across the literature, and they are not always consistent.
For example, the original definitions of Type-3 and Type-4 clones by \citet{Carter1993} differ significantly from the definitions by \citet{Roy2009} often cited in clone detection work:

\medskip\noindent A Type-3 clone is
\begin{itemize}
\item A clone with very similar source code, but with small changes made to the code to tailor it to some new function~\citep{Carter1993}.
\item A copied fragment with further modifications such as changed, added, or removed statements, in addition to variations in identifiers, literals, types, whitespace, layout, and comments~\citep{Roy2009}.
\end{itemize}
A Type-4 clone is
\begin{itemize}
\item A functionally identical clone, possibly with the originator unaware that there is a function already available that accomplishes essentially the same function~\citep{Carter1993}.
\item Two or more code fragments that perform the same computation but are implemented by different syntactic variants~\citep{Roy2009}.
\end{itemize}
The variation and ambiguity in the definitions show that there is no generally accepted definition of clones or their classification, and the definitions are subject to interpretation, causing confusion in the community.

Note that Type-4 clones, as defined by \citet{Carter1993} or by \citet{Roy2009}, require \emph{identical} functionality or the same computation.
According to Rice's theorem, the above definitions are not decidable in general, and similar to the definition of clones, there is no generally accepted definition of semantic clones.\footnote{The first paper on semantic clones~\citep{Gabel2008} defines semantic clones as having isomorphic subgraphs in program dependence graphs~\citep{Krinke2001}.}
\changed{However, work on semantic clone detection also uses a broader criterion based on similar or equal run-time behaviour rather than requiring identical functionality~\citep{Thaller2022}.}
Moreover, the original definition of \citet{Carter1993} does not require that Type-4 clones are implemented by different syntactic variants, while the definition of \citet{Roy2009} does.
Requiring syntactic dissimilarity is problematic because it is not clear \emph{how much} syntactic similarity is allowed for two code fragments to be considered Type-4 and not Type-3 clones, as the definitions of Type-3 clones by \citet{Roy2009} and \citet{Svajlenko2014} do not limit the number of added or deleted statements.

\section{BigCloneBench}

BigCloneBench is a dataset consisting of method-level clones in Java programs, i.e., the fragments that are considered for clone detection are complete methods.
\changed{For each retrieved method receiving a decision, BigCloneBench assigns a method-level true-positive label if it has the specified functionality or a method-level false-positive label if it does not.}
BigCloneBench is available in multiple versions.
An initial version~\citep{Svajlenko2014} only contained clones derived from 10 functionalities, 42 exemplar functions, 6,000 methods tagged as true positives, and 53,688 methods tagged as false positives, leading to 6,164,953 true clone pairs and 258,574 false clone pairs.
There exist at least two released versions of the earlier version of BigCloneBench; however, the released versions are slightly different from the reported numbers above.
The second version of the benchmark was introduced in 2016 as a part of BigCloneEval~\citep{Svajlenko2015,Svajlenko2016}, a clone detection tool evaluation framework with BigCloneBench. 
It contains 43 specified functionalities with a total of 7,868,560 true clone pairs~\citep{Svajlenko2015}, later expanded to 8,375,313 true clone pairs~\citep{Svajlenko2016}.
The most detailed explanation is available in the thesis by \citet{Svajlenko2017}, where the size is given as 8,915,130 true clone pairs and 288,367 false clone pairs.

The discussion in this paper mainly uses the reported numbers~\citep{Svajlenko2017} and will sometimes use numbers extracted from the released version of BigCloneBench, which has been released individually\footnote{\url{https://github.com/clonebench/BigCloneBench}} and as a part of the evaluation framework\footnote{\url{https://github.com/jeffsvajlenko/BigCloneEval}}.
The downloaded version contains 8,584,153 true clone pairs, 279,032 false clone pairs, and 43 specified functionalities.
\changed{The reported counts differ from those measured in the downloaded version.}

\subsection{Construction of BigCloneBench}

To understand the features of BigCloneBench that affect the ground truth quality, it is necessary to understand how BigCloneBench has been constructed.
At the core of the construction is the validation of methods by human judges:
\emph{Instead of asking the judges to validate if two methods are similar, they have been asked to validate if an individual function implements a target functionality.
Clones are identified as methods that share functionality}~\citep{Svajlenko2017}.
The authors of BigCloneBench chose this approach to reduce subjectivity in judging whether two methods are clones of each other.
Note that the identification only requires the methods to share functionality without specifying the amount of shared functionality, potentially leading to pairs of methods that are not clones being labelled as true clone pairs because they share a minor functionality while their main overall functionality is very different.
However, the Type-4 definition they presented is \emph{syntactically dissimilar code fragments that implement the same functionality}~\citep{Svajlenko2014,Svajlenko2015,Svajlenko2016}, which is a much stronger requirement than the one used in the construction of BigCloneBench.

\begin{figure}[tbp]
    \centering
    \includegraphics[width=0.7\textwidth]{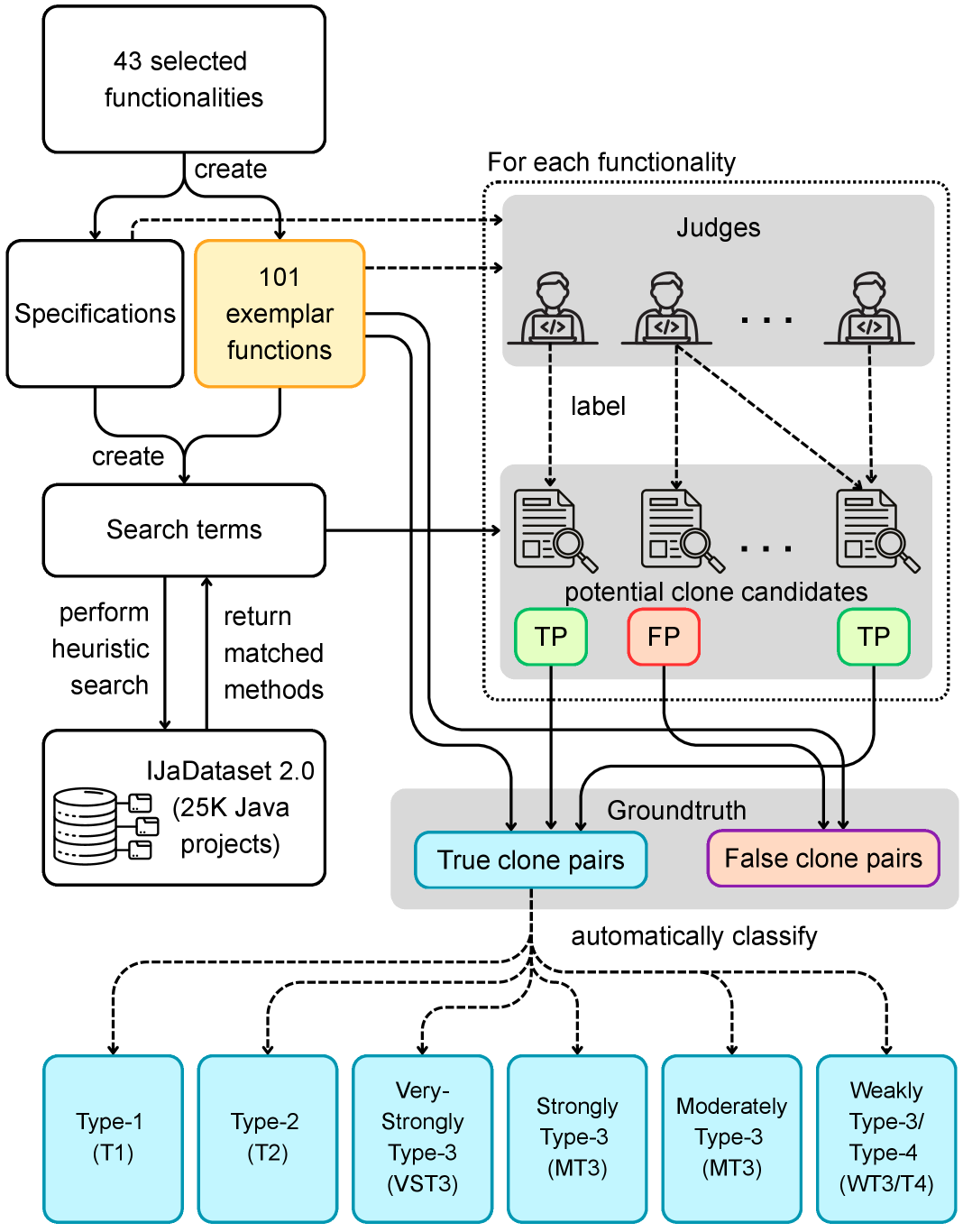} 
    \caption{The process of constructing BigCloneBench.}
    \label{fig:bcb_construction} 
\end{figure}

We present the construction process of BigCloneBench in a simplified way, as depicted in Figure~\ref{fig:bcb_construction}, in the following.
 
\subsubsection{Source Code}

BigCloneBench is built from the IJaDataset 2.0, a dataset of 250M LOC 
in 2.5M Java files from 25K projects mined from SourceForge and Google Code~\citep{Svajlenko2017}.
\subsubsection{Functionalities}
Functionalities are the main building blocks of BigCloneBench. They are defined as tasks that are expected to appear often in software development, such as ``Copy File'', ``Connect to FTP Server'', or ``Download From Web''.
A set of functionalities that are commonly needed in open source Java software has been identified, and 43 functionalities have been selected for BigCloneBench.

\subsubsection{Specifications}

For each of the 43 functionalities, a specification has been created that describes the functionality.
The specifications are kept simple and are not intended to be exhaustive, and they usually do not contain any details.
For example, the specification for the ``Copy File'' functionality is simply ``copies a file''.

\subsubsection{Exemplar Functions}
The core of BigCloneBench is a set of 101 exemplar functions, which have been composed as example implementations of 43 selected functionalities that are expected to appear often
(the downloaded version contains 107 exemplar functions from which 6 do not appear in any pair).
22 of the 43 selected functionalities have a single exemplar function.
In earlier papers about BigCloneBench, these exemplar functions were termed ``sample snippets''; however, in this paper, we use the clearer term ``exemplar functions'' as it has been introduced in newer publications on BigCloneBench~\citep{Svajlenko2021}.

\begin{figure*}[tbh]
	\footnotesize
	\begin{verbatim}
public static void copyFile2(File srcFile, File destFile) throws IOException {
  FileUtils.copyFile(srcFile, destFile);
}
\end{verbatim}
\caption{Copy File Exemplar Function (Method 23677115 in file \texttt{CopyFileSamples.java}, lines 38--40).}
\label{fig:copyFile}
\end{figure*}
		
An example of a selected functionality is ``Copy File'' and a corresponding exemplar function is shown in \Cref{fig:copyFile}.
For the ``Copy File'' functionality, six exemplar functions have been created.

\subsubsection{Search Terms for Potential Clone Candidates}

From the exemplar functions and the specification of the 43 functionalities, a heuristic search has been performed to find methods that use the 43 functionalities.
The heuristic search has resulted in 77,933 methods, which are the candidates for potential clones within each of the 43 functionalities.
A potential clone candidate will later become a confirmed clone if the candidate is paired with another candidate in a true clone pair.

\begin{figure}
	\footnotesize
	\begin{verbatim}
		[getChannel] OR [transferFrom] 
		OR [FileUtils.copyFile] OR [read AND write]
		OR [nextLine AND [print OR println OR write] 
		    OR [IOUtils.copy] OR [IOUtils.copyLarge]
	\end{verbatim}
	\caption{Search terms for the ``Copy File'' functionality.}
	\label{fig:heuristic}
\end{figure}

For example, \Cref{fig:heuristic} shows the search terms to identify methods that contain the ``Copy File'' functionality, extracted from the downloaded dataset.
Note that the search is not syntactically correct as given.
37,102 methods matched the search heuristic for the ``Copy File'' functionality and are candidates for potential clones.

\subsubsection{Labelling of Potential Clone Candidates}

A set of judges then compared the candidate methods of the 43 functionalities to the specifications and exemplar functions of a functionality and labelled them as true positives or false positives.
Overall, 15,290 methods have been labelled as true positives and 61,756 methods have been labelled as false positives.
Only a small number of methods (9,533 -- 12\%) have been labelled by more than one judge, and the final label for them is the majority vote (887 methods with an equal number of votes are labelled as undecided).
This means that 68,400 methods (88\%) have been labelled by a single judge only.

The released dataset has slightly different numbers and contains 75,673 retrieved methods.
Some of the methods are duplicated in the dataset.
Of the retrieved methods, 73,906 are unique.
Moreover, from the 14,891 methods labelled as true positives, 14,679 are unique, and from the 60,782 methods labelled as false positives, 60,019 are unique.

\changed{The search heuristics were designed to identify as many true-positive methods as possible without overburdening the judges in their tagging efforts.}
However, instructions to the judges were set in a way that allowed that target functionality to be only a part of the method:
\textit{``True positives may exceed the specification by performing additional related or unrelated tasks.''~\citep{Svajlenko2017}.}
This even allowed methods to be tagged as true positives when the target functionality is only a small part of the method.
\changed{While it is widely accepted that empirical code similarity research must slightly relax the strict theoretical requirement of mathematically identical functionality for Type-4 clones, the methodology of BigCloneBench relaxes this constraint too far by allowing methods that only share a minor sub-task to be labelled as true semantic clones.}
\changed{Broad candidate retrieval is not itself problematic and resembles the coverage objective of an oracle for automated precision evaluation, where omitted true clones may otherwise be counted as false detections.}
\changed{However, BigCloneBench remains too incomplete to serve as such a precision oracle, and its permissive acceptance criterion is unsuitable for constructing reliable positive reference pairs for recall evaluation of same-main-function similarity when the resulting pairs are not subsequently validated.}

Because a method can have more than one of the 43 functionalities, 209 methods are true positives for more than one functionality.
For example, method 10151252 has been labelled as a true positive for the ``Copy File'' functionality because the method copies (uploads) a file to an FTP server, and also as a true positive for the ``Connect to FTP Server'' functionality.

\subsubsection{Ground Truth of Clone Pairs}

The ground truth\footnote{Technically, the BigCloneBench authors never use the term.  However, as we focus on the use of BigCloneBench in machine learning where BigCloneBench is often used as the ground truth, we use the term ground truth for the labelled data in BigCloneBench.} is constructed from the exemplar functions and the potential clone candidates.
The exemplar functions are in sets $X_f$ for each of the 43 functionalities ($f$).
\changed{For each of the 43 functionalities $f$, methods with a method-level true-positive label are in set $P_f$, while methods with a method-level false-positive label are in set $N_f$.}

The ground truth of true and false clone pairs is constructed per functionality $f$, i.e., for pairs of methods $(m, n)_f$.
The ground truth is constructed as follows:
\changed{The pairs $(m, n)_f$ and $(n, m)_f$ are labelled as true clone pairs if $(m \in X_f \cup P_f) \land (n \in X_f \cup P_f)$, that is, if both methods are exemplar functions or carry method-level true-positive labels.}
\changed{The pairs $(m, n)_f$ and $(n, m)_f$ are labelled as false clone pairs if $(m \in X_f) \land (n \in N_f)$ for a functionality $f$.}
\changed{False clone pairs are therefore constructed only between an exemplar function and a candidate method carrying a method-level false-positive label.}
\changed{Pairs of two candidate methods carrying that label remain unlabelled because they may still be clones of each other.}
\changed{The automatic cross-product therefore converts method-level evidence that two methods each contain a target functionality into a pair-level assertion that they are clones.}
\changed{Together with the permissive method-labelling criterion, this creates the central construction flaw for same-main-function similarity: a benchmark intended to evaluate recall contains positive reference pairs whose methods may differ completely in their \textbf{major} functionality and share only a \textbf{minor} functionality.}
\changed{The result is a flawed WT3/T4 ground truth under the usual definition of Type-4 clones, as a detector that correctly rejects such a pair is counted as failing to detect a supposed clone.}
\changed{The downloaded true-pair table distinguishes \texttt{sample-sample} pairs between two exemplar functions, \texttt{sample-tagged} pairs between an exemplar function and a method tagged as a true positive, and \texttt{tagged-tagged} pairs between two methods tagged as true positives.}
\changed{It contains 206, 42,175, and 8,541,772 true pairs in these categories, respectively. Applying the documented WT3/T4 selection criterion leaves 49, 40,697, and 8,381,611 pairs.}

\changed{Using the reported construction counts for the running example of the ``Copy File'' functionality, 6 exemplar functions and 3,084 true-positive methods give 4,772,505 possible true clone pairs, and 6 exemplar functions and 34,018 false-positive methods give 204,108 possible false clone pairs.}

The above construction leads to a ground truth of 8,915,130 true clone pairs and 288,367 false clone pairs.
\changed{None of the true or false clone pairs have been manually validated, and the authors of BigCloneBench have not provided any information on the quality of the ground truth of clone pairs.}

\subsubsection{Automatic Classification}
In the last step of the construction, each true clone pair is classified as Type-1 (T1), Type-2 (T2), Very-Strongly Type-3 (VST3), Strongly Type-3 (ST3), Moderately Type-3 (MT3), and Weakly Type-3/Type-4 (WT3/T4).
For the comparison, all methods are normalised by removing all comments and pretty printing.
Type-1 pairs are pairs that only differ in their whitespace.
For Type-2, the methods are identical after all identifiers are replaced by a common identifier and all literal values are replaced by a common literal value.
For all other pairs, the similarity of the two methods is measured as the minimum ratio of the lines or tokens one method shares with the other after normalisation.
If the similarity is below 0.5, the pair is classified as a Weakly Type-3/Type-4 clone pair. Otherwise, the pair is classified as a Very Strongly Type-3 ($\ge$0.9), Strongly Type-3 clone pair ($\ge$0.7) or Moderately Type-3 clone pair ($\ge$0.5).
These thresholds have not been tied to prior benchmarks, human judgements, or sensitivity analysis \citep{Svajlenko2014}.

\changed{8,498,894 out of 8,915,130 pairs have been classified as Weakly Type-3/Type-4 (95.3\%), leaving only 416,236 pairs (4.7\%) across all other clone categories combined.}
\changed{There has not been any further distinction within the WT3/T4 category because the automatic classification relies solely on textual similarity thresholds, without manual validation or semantic analysis to establish which pairs are indeed functionally similar.}

\subsection{Usage of BigCloneBench}
\paragraph*{Precision and Recall:~~} 

\changed{BigCloneBench was designed to measure the recall of clone detection tools.}
\changed{Reliable recall evaluation requires the positive reference pairs to be genuine clones under the clone definition being evaluated, because a false reference positive penalises a detector that correctly does not report that pair.}
BigCloneBench has not been intended for and should not be used for directly measuring precision.
\changed{The reason is that, for most method pairs, no ground truth is available, and it is unknown whether the pair is a clone or a non-clone.}
Instead, precision needs to be evaluated differently, by manually investigating a representative sample of the results of the clone detection or code similarity detection.
With the available ground truth data, only the lower and upper bounds of precision can be determined, which can be used for precision estimation.
\changed{Thus, although the candidate search pursued the coverage objective relevant to a precision oracle, the resulting ground truth is too incomplete for direct precision evaluation, while its permissive method labels undermine the reliable positive reference set required for recall evaluation of same-main-function similarity.}

\paragraph*{BigCloneBench for Semantic Clone Detection:~~}

\changed{The dataset was not intended for semantic clone detection~\citep{Svajlenko2022}.}
Instead, the automatic classification should be seen as resulting in pairs that share some (minor) functionality and are textually similar to a certain degree.

\section{Experiments to Assess the Ground Truth Quality}

\changed{Section~3 identifies a mechanism by which methods that share only a subordinate functionality can be paired automatically as true clones, but the construction record alone does not establish how often this occurs.}
\changed{Our earlier study of pairs from one functionality~\citep{Krinke2022} provided an initial indication that the effect could be substantial, but did not constitute a systematic assessment of WT3/T4 ground-truth quality.}

\begin{figure}[tbp]
\centering
\includegraphics[width=\textwidth]{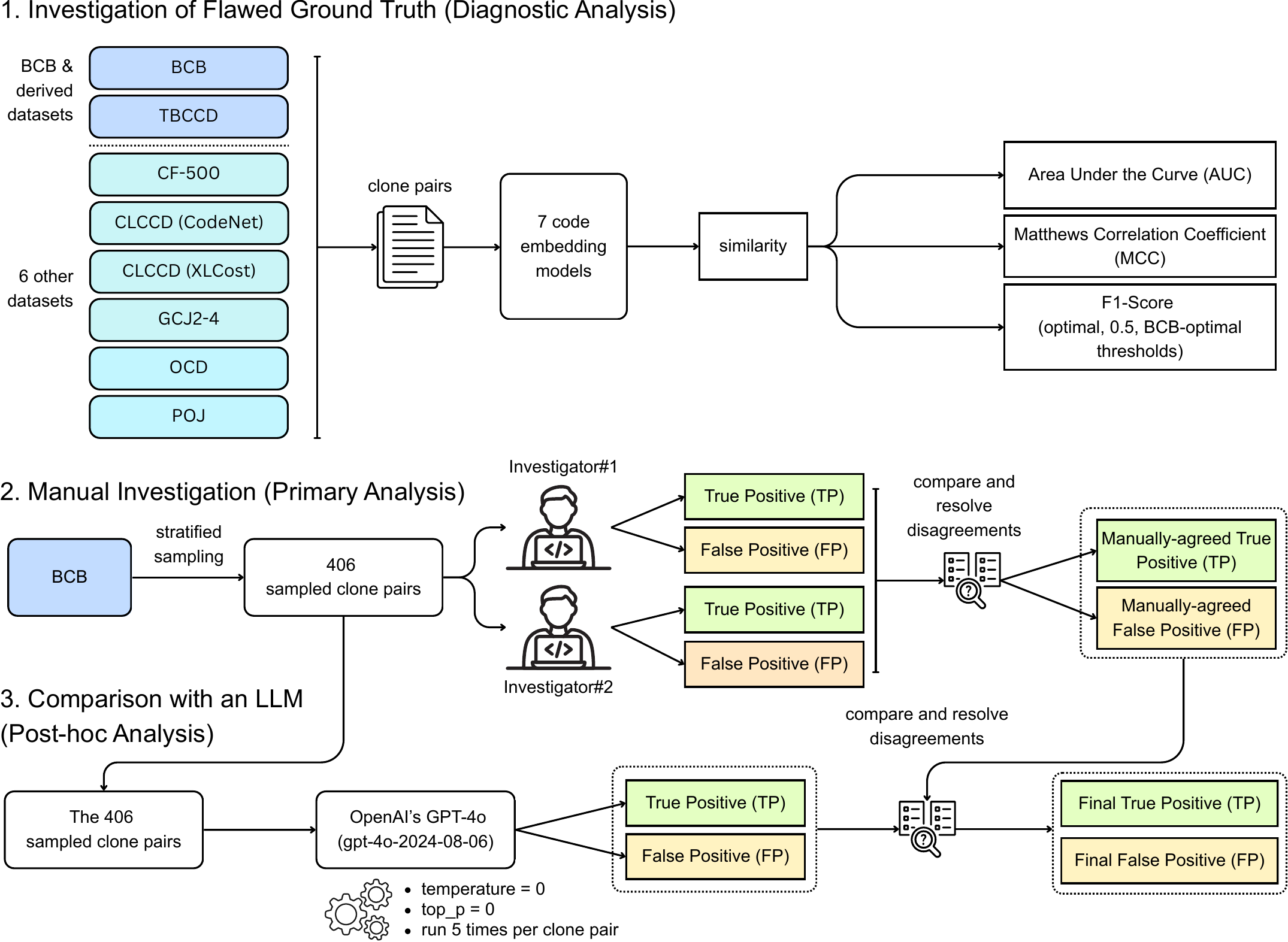}
\caption{\changed{Overview of the experiments to assess the ground-truth quality of BigCloneBench.}}
\label{fig:experiments}
\end{figure}

\begin{table}[tbp]
\centering
\footnotesize
\setlength{\tabcolsep}{3pt}
\begin{tabular}{@{}>{\raggedright\arraybackslash}p{0.12\textwidth}>{\raggedright\arraybackslash}p{0.22\textwidth}>{\raggedright\arraybackslash}p{0.34\textwidth}>{\raggedright\arraybackslash}p{0.22\textwidth}@{}}
\toprule
\changed{Step} & \changed{Input} & \changed{Role and output} & \changed{Limitation} \\
\midrule
\changed{Diagnostic} & \changed{BigCloneBench, TBCCD, and six comparison datasets} & \changed{Scalable indication of an anomalous embedding-score pattern} & \changed{Heuristic, with no population estimate} \\
\changed{Estimate} & \changed{A 406-pair stratified probability sample} & \changed{Population-weighted estimate and analysis of the construction mechanism} & \changed{Definition-dependent, with a conservative variance bound for singleton strata} \\
\changed{Sensitivity} & \changed{The 406 manually labelled pairs and human consensus labels} & \changed{Post-hoc challenge using LLM labels and explanations} & \changed{Not an independent oracle} \\
\changed{Literature} & \changed{179 papers using BigCloneBench} & \changed{Estimated number of papers with potentially affected results} & \changed{Aggregate validity threat, not a verdict on each paper} \\
\changed{Synthesis} & \changed{The empirical components and further dataset observations} & \changed{Implications and guidance on alternative benchmarks} & \changed{Task-specific guidance, not an audit of every alternative} \\
\bottomrule
\end{tabular}
\caption{\changed{Evidence chain connecting the components of the investigation to their roles and the limitations on the resulting claims.}}
\label{tab:evidence_chain}
\end{table}

\changed{\Cref{fig:experiments} shows the procedures for the three ground-truth assessments, while \Cref{tab:evidence_chain} summarises their roles and limitations together with those of the later literature study and synthesis.}
		
\subsection{\changed{Exploratory Embedding Analysis}}

\changed{The first assessment uses seven contemporary, publicly available embedding models as a scalable diagnostic, comparing full and balanced BigCloneBench, whose positive pairs are dominated by WT3/T4, and TBCCD with six datasets curated for functional similarity: CF-500~\citep{Xue2022}, CLCCD (CodeNet)~\citep{Moumoula2025}, CLCCD (XLCoST)~\citep{Moumoula2025}, GCJ2-4 (Full)~\citep{Li2023}, OCD\footnote{\changed{For OCD, we first excluded the 100 self-pairs and retained only one orientation of each pair, leaving 450 true clone pairs and 4,500 false clone pairs.}}~\citep{Ragkhitwetsagul2018}, and POJ~\citep{Li2017}.}
\changed{The details of the datasets are discussed in \Cref{sec:datasets}.}

\changed{Initially, the experiment uses \emph{jina-code-embeddings-1.5B}, selected because it is specifically fine-tuned for source code and reports strong performance on code-retrieval tasks~\citep{kryvosheieva2025efficient}.}
\changed{The six additional models were selected to test robustness across code-specific and general-purpose embedding models, Qwen- and Mistral-derived bases, and parameter scales from 1.5B to 8B.}
\changed{The resulting set covers these dimensions but is not exhaustive.}
\changed{The documented contrastive-training datasets do not include BigCloneBench or any other benchmark evaluated here, although they include other Codeforces data. However, the model inherits a Qwen2.5-Coder backbone pre-trained on a broader public-code corpus, so exposure to the underlying source code cannot be excluded~\citep{kryvosheieva2025efficient,Hui2024Qwen}.}

\changed{Because the models are heuristic measures of functional similarity, this comparison can reveal anomalous agreement patterns but cannot establish the prevalence of WT3/T4 mislabelling.}
\changed{The six comparison datasets provide a sanity check: across all seven embedding models, agreement with their ground truths is consistently higher than for balanced BigCloneBench and TBCCD.}

\begin{table*}
\caption{Results (AUC) of the experiment measuring the correlation between the ground truth and the functional similarity as determined by jina-code-embeddings-1.5B for functional similarity.
$^*$ indicates results that are above 0.9.}
\label{tab:resultsAUC}
\centering
\begin{tabular}{l|r|r}
\toprule
Dataset & \multicolumn{1}{c|}{\# Pairs} & AUC \\
\midrule
BigCloneBench -- BCB (balanced) & 558,064 & 0.75 \\
BigCloneBench -- BCB (full) & 8,863,185 & 0.75 \\
TBCCD~\citep{Yu2019} & 672,996 & 0.79 \\
\midrule
CF-500~\citep{Xue2022} & 1,550,154 & $^*$1.00 \\
CLCCD (CodeNet)~\citep{Moumoula2025} & 6,000 & $^*$1.00 \\
CLCCD (XLCoST)~\citep{Moumoula2025} & 6,000 & $^*$1.00 \\
GCJ2-4 (Full)~\citep{Li2023} & 70,355,828 & $^*$0.99 \\
OCD~\citep{Ragkhitwetsagul2018} & 900 & $^*$1.00 \\
POJ~\citep{Li2017} & 25,948,000 & $^*$1.00 \\
\bottomrule
\end{tabular}
\end{table*}

\begin{table*}
\caption{Results (MCC and F1) of the experiment for the individual optimal thresholds and for a similarity threshold of 0.5.
$^*$ indicates results that are above 0.9.}
\label{tab:results}
\centering
\begin{tabular}{l|c@{~~}r|c@{~~}r|rr}
\toprule
Dataset & Opt.~Thres. & MCC & Opt.~Thres. & F1 & MCC@0.5 & F1@0.5 \\
\midrule
BCB (balanced) & 0.19 & 0.38 & 0.17 & 0.73 & 0.23 & 0.29 \\
BCB (full) & 0.14 & 0.19 & 0.00 & 0.98 & 0.07 & 0.30 \\
TBCCD & 0.24 & 0.44 & 0.19 & 0.73 & 0.27 & 0.27 \\
\midrule
CF-500 & 0.39 & $^*$0.99 & 0.39 & $^*$0.99 & $^*$0.97 & $^*$0.99 \\
CLCCD (CodeNet) & 0.26 & $^*$0.99 & 0.26 & $^*$1.00 & $^*$0.93 & $^*$0.96 \\
CLCCD (XLCoST) & 0.36 & $^*$1.00 & 0.36 & $^*$1.00 & $^*$1.00 & $^*$1.00 \\
GCJ2-4 (Full) & 0.53 & $^*$0.94 & 0.52 & $^*$0.97 & $^*$0.94 & $^*$0.97 \\
OCD & 0.53–0.79 & $^*$1.00 & 0.53–0.79 & $^*$1.00 & $^*$1.00 & $^*$1.00 \\
POJ & 0.44 & $^*$0.97 & 0.44 & $^*$0.98 & $^*$0.96 & $^*$0.98 \\
\bottomrule
\end{tabular}
\end{table*}

For each of the datasets, we extracted the source code fragments and the labelled pairs.
To demonstrate the effect of the imbalance in the BigCloneBench ground truth, we included the full BigCloneBench dataset, which has 8,584,153 true clone pairs, but only 279,032 false clone pairs (3.1\%).
A predictor that would label all pairs as true clone pairs would achieve a perfect recall of 100\% and a high precision of 96.9\%, leading to an F1 score of 98.4\%.
This means that the predictor would report 8,584,153 + 279,032 = 8,863,185 total pairs as true clone pairs, of which 8,584,153 are true positives and 279,032 are false positives, with no true or false negatives, leading to $\textrm{F1} = \frac{2\mathrm{TP}}{2\mathrm{TP} + \mathrm{FP}+\mathrm{FN}} \approx 0.984$.
We therefore created a balanced version of BigCloneBench, where we only kept 279,032 randomly selected true clone pairs and the 279,032 false clone pairs.
For other datasets, we only used the balanced versions, i.e., the number of true pairs and false pairs is the same.
For datasets that were not already balanced, we applied random under-sampling of the majority class to achieve balance (similar to the approach used for the balanced version of BigCloneBench).

The chosen model is then used to compute embeddings for all fragments of all datasets.
We then calculated the cosine similarity for each of the complete sets, covering a total of 107,961,127 pairs, and computed the following metrics:
\begin{itemize}
\item Area Under the Curve (AUC) of the Receiver Operating Characteristic (ROC) curve.
\item \changed{Matthews} Correlation Coefficient (MCC) and F1-score at two thresholds: the optimal threshold and the threshold 0.5.
\end{itemize}
\changed{AUC is a threshold-independent metric that summarises the model's ability to distinguish labelled clone pairs from labelled non-clone pairs across all possible thresholds.}
\changed{MCC and F1-score are threshold-dependent metrics that require setting a threshold to predict whether each pair is a clone or a non-clone.}
We report both the MCC and F1-score despite previous work on code similarity reporting only F1 scores and not MCC scores, because MCC is a more reliable metric for binary classification, especially for imbalanced datasets~\citep{Chicco2023,Yao2020assessing}.
For example, the same predictor labelling all pairs as true clone pairs would report an MCC of 0\% for the full imbalanced BigCloneBench ground truth.

The results of the experiment are shown in \Cref{tab:resultsAUC} for AUC and in \Cref{tab:results} for F1 and MCC.
The results show that the six curated datasets achieve near-perfect results with AUC, MCC, and F1-score close to 1.0 (all scores are above 0.9).
This indicates a strong correlation between the ground truth and the functional similarity as measured by the embedding model.
\changed{In contrast, the balanced BigCloneBench and TBCCD achieve substantially lower scores, indicating a weaker correlation between the ground truth and the functional similarity as measured by the embedding model.}
\changed{Both versions of BigCloneBench achieve an AUC of 0.75, and TBCCD achieves an AUC of 0.79, indicating that the embedding model can distinguish labelled clone pairs from labelled non-clone pairs to some extent, but not as well as for the curated datasets, which have AUCs ranging from 0.99 to 1.00.}
\changed{These comparisons are descriptive: no pair-level significance test was performed because methods recur across pairs and the observations are therefore not independent.}

\changed{The full imbalanced BigCloneBench, however, achieves a much lower MCC of 0.19 at its optimal MCC threshold of 0.14 and an MCC of 0.07 at a threshold of 0.5, very close to random.}
At a threshold of 0.5, the reported F1-score is 0.30, almost the same as for the balanced dataset.
However, the above-discussed problem of using F1 scores for imbalanced datasets can be clearly seen in the results:
A predictor that labels all pairs as true clone pairs would achieve an F1 score of 0.98.
Consequently, the optimisation of the threshold is down to 0, yielding indeed a highly misleading F1-score of 0.98.
This highlights the importance of using a more reliable metric, such as MCC, for imbalanced datasets, as the F1-score can be misleadingly high even when the model is performing poorly.
\changed{Moreover, it highlights that the full BigCloneBench should not be used as ground truth for training or evaluating same-main-function semantic similarity, because the reported precision, recall, and F1-score may appear very high even when the approach is performing poorly at that task.}

\changed{The optimal MCC thresholds for the balanced BigCloneBench (0.19) and TBCCD (0.24) are also lower than those of the curated datasets, which range from 0.26 to 0.79, including OCD's optimal interval of 0.53--0.79.}
The MCC scores at the optimal thresholds for the balanced BigCloneBench (0.38) and TBCCD (0.44) are also much lower than those of the curated datasets (ranging from 0.94 to 1.00), indicating a weaker correlation between the ground truth and the functional similarity as measured by the embedding model.

\changed{Balanced BigCloneBench and TBCCD are clear outliers in the above experiment, with substantially lower scores across all metrics than the other benchmarks.}
\changed{This contrast is compatible with misalignment between the WT3/T4 labels and functional similarity as measured by the embedding models, but the diagnostic alone cannot distinguish this from alternative explanations, including that BigCloneBench presents a harder problem or operationalises a different construct.}
\changed{For illustration, if the embedding model were perfect (100\% recall and precision) and only true pairs were mislabelled, the observed scores could be converted into implied proportions of mislabelled true pairs.}
\changed{For balanced BigCloneBench at the threshold of 0.19, an MCC of 0.38 implies that around 75\% of true labels are wrong under these assumptions:}
$\mathit{TP} = \frac{1-x}{2}$, $\mathit{TN} = \frac{1}{2}$, $\mathit{FN} = \frac{x}{2}$, $\mathit{FP} = 0$, $\textrm{MCC} = \frac{\mathit{TP} \cdot \mathit{TN} - \mathit{FP} \cdot \mathit{FN}}{\sqrt{(\mathit{TP} + \mathit{FP})(\mathit{TP} + \mathit{FN})(\mathit{TN} + \mathit{FP})(\mathit{TN} + \mathit{FN})}} = \frac{(1-x)}{\sqrt{(1+x)(1-x)}}$, $\textrm{MCC} = 0.38$ for $x = 0.7476$.
\changed{At the threshold of 0.5, an MCC of 0.23 increases the implied proportion to around 90\%.}
\changed{Similarly, an F1 score of 0.73 at the threshold of 0.17 implies around 43\%, while an F1 score of 0.29 at the threshold of 0.5 implies around 83\%.}
\changed{These values illustrate the implications of the stated assumptions. They are not estimates of the actual prevalence of mislabelling, which is instead estimated by the manual investigation.}

\begin{quote}
	\changed{\textbf{Across all seven embedding models, balanced BigCloneBench and TBCCD have lower agreement between embedding similarity and ground truth than the six comparison datasets. This provides a scalable diagnostic of potential WT3/T4 mislabelling, not a population estimate.}}
\end{quote}

\begin{table*}
\caption{Results of the experiment when using the optimal thresholds as computed for BigCloneBench.
$^*$ indicates results that are above 0.9.}
\label{tab:resultsBCB}
\centering
\begin{tabular}{l|rr}
\toprule
Dataset & MCC@0.19 & F1@0.17 \\
\midrule
BigCloneBench & – & – \\
TBCCD & 0.41 & 0.73 \\
\midrule
CF-500 & $^*$0.91 & $^*$0.94 \\
CLCCD (CodeNet) & $^*$0.98 & $^*$0.98 \\
CLCCD (XLCoST) & $^*$0.94 & $^*$0.95 \\
GCJ2-4 (Full) & 0.31 & 0.70 \\
OCD & 0.36 & 0.71\\
POJ & 0.79 & 0.88 \\
\bottomrule
\end{tabular}
\end{table*}

\changed{At the thresholds optimised for balanced BigCloneBench (0.19 for MCC and 0.17 for F1), most other datasets score below their own optima, with the largest reductions for GCJ2-4 (Full) and OCD (\Cref{tab:resultsBCB}).}
\changed{The transferred thresholds nevertheless retain MCC and F1 scores above 0.9 for CF-500 and both CLCCD datasets, so their effectiveness varies by dataset.}
\changed{In contrast, at a threshold of 0.5, which lies within the range of optimal thresholds of the six curated datasets (0.26--0.79), their MCC and F1 scores remain high (0.93--1.00).}
\changed{At this threshold, the MCC and F1 scores for balanced BigCloneBench and TBCCD range from 0.23 to 0.29, while full BigCloneBench has an MCC of 0.07 and an F1 score of 0.30.}

\changed{To assess whether the results are specific to the initially chosen model, \emph{jina-code-embeddings-1.5B}, we repeated the experiment with the other six models.}
\Cref{tab:embeddingmodels} shows the results for AUC, F1, and MCC for all models and all datasets.
\changed{The results show that the six curated datasets achieve higher AUC, F1, and MCC scores than balanced BigCloneBench and TBCCD with all seven embedding models, showing that the observed contrast is not confined to the initially chosen model.}

\begin{table}
\caption{Performance comparison of code embedding models across different benchmarks.
\changed{All datasets are balanced, including BigCloneBench.}
\changed{F1 and MCC use thresholds optimised separately for each metric, model, and dataset.}
\changed{Within each benchmark, the highest values for each metric are shown in bold and the lowest values are underlined, except when all models have the same value.}}
\label{tab:embeddingmodels}
\renewcommand{\arraystretch}{0.82}
\begin{tabular}{llrrr}
\toprule
Benchmark & Model & AUC & F1 & MCC \\
\midrule
BigCloneBench
& C2LLM-7B & 0.776 & 0.741 & 0.420 \\
& jina-code-embeddings-1.5b & 0.749 & 0.728 & 0.380 \\
& Linq-Embed-Mistral & \textbf{0.814} & \textbf{0.778} & \textbf{0.510} \\
& Octen-Embedding-8B & 0.740 & 0.725 & 0.365 \\
& Qodo-Embed-1-7B & \underline{0.726} & \underline{0.702} & \underline{0.327} \\
& Qwen3-Embedding-8B & 0.737 & 0.722 & 0.360 \\
& SFR-Embedding-2\_R & 0.803 & 0.766 & 0.481 \\
\midrule
TBCCD
& C2LLM-7B & \underline{0.770} & 0.730 & 0.406 \\
& jina-code-embeddings-1.5b & 0.794 & 0.734 & 0.442 \\
& Linq-Embed-Mistral & 0.783 & 0.727 & 0.407 \\
& Octen-Embedding-8B & 0.795 & 0.742 & 0.445 \\
& Qodo-Embed-1-7B & \textbf{0.810} & \textbf{0.749} & \textbf{0.461} \\
& Qwen3-Embedding-8B & 0.774 & 0.723 & 0.405 \\
& SFR-Embedding-2\_R & 0.778 & \underline{0.722} & \underline{0.399} \\
\midrule
CF500
& C2LLM-7B & 0.998 & 0.984 & 0.968 \\
& jina-code-embeddings-1.5b & \textbf{0.999} & \textbf{0.993} & \textbf{0.985} \\
& Linq-Embed-Mistral & 0.986 & 0.943 & 0.889 \\
& Octen-Embedding-8B & \textbf{0.999} & 0.989 & 0.978 \\
& Qodo-Embed-1-7B & 0.990 & 0.952 & 0.904 \\
& Qwen3-Embedding-8B & \textbf{0.999} & 0.992 & 0.984 \\
& SFR-Embedding-2\_R & \underline{0.984} & \underline{0.941} & \underline{0.883} \\
\midrule
CLCCD (CodeNet)
& C2LLM-7B & \textbf{1.000} & 0.992 & 0.984 \\
& jina-code-embeddings-1.5b & \textbf{1.000} & 0.995 & 0.991 \\
& Linq-Embed-Mistral & \underline{0.997} & \underline{0.979} & \underline{0.957} \\
& Octen-Embedding-8B & \textbf{1.000} & \textbf{0.998} & \textbf{0.997} \\
& Qodo-Embed-1-7B & \underline{0.997} & 0.983 & 0.966 \\
& Qwen3-Embedding-8B & \textbf{1.000} & 0.996 & 0.992 \\
& SFR-Embedding-2\_R & \underline{0.997} & 0.980 & 0.959 \\
\midrule
CLCCD (XLCoST) 
& C2LLM-7B & \textbf{1.000} & 0.999 & 0.999 \\
& jina-code-embeddings-1.5b & \textbf{1.000} & 0.999 & 0.998 \\
& Linq-Embed-Mistral & \textbf{1.000} & \textbf{1.000} & \textbf{1.000} \\
& Octen-Embedding-8B & \textbf{1.000} & \textbf{1.000} & \textbf{1.000} \\
& Qodo-Embed-1-7B & \underline{0.999} & \underline{0.998} & \underline{0.996} \\
& Qwen3-Embedding-8B & \textbf{1.000} & 0.999 & 0.999 \\
& SFR-Embedding-2\_R & \textbf{1.000} & \textbf{1.000} & 0.999 \\
\midrule
GCJ2-4 (Full)
& C2LLM-7B & 0.990 & 0.976 & 0.953 \\
& jina-code-embeddings-1.5b & 0.986 & 0.970 & 0.940 \\
& Linq-Embed-Mistral & 0.937 & 0.868 & 0.743 \\
& Octen-Embedding-8B & \textbf{0.991} & \textbf{0.983} & \textbf{0.967} \\
& Qodo-Embed-1-7B & 0.966 & 0.914 & 0.830 \\
& Qwen3-Embedding-8B & 0.971 & 0.928 & 0.855 \\
& SFR-Embedding-2\_R & \underline{0.936} & \underline{0.867} & \underline{0.741} \\
\midrule
OCD
& C2LLM-7B & 1.000 & 1.000 & 1.000 \\
& jina-code-embeddings-1.5b & 1.000 & 1.000 & 1.000 \\
& Linq-Embed-Mistral & 1.000 & 1.000 & 1.000 \\
& Octen-Embedding-8B & 1.000 & 1.000 & 1.000 \\
& Qodo-Embed-1-7B & 1.000 & 1.000 & 1.000 \\
& Qwen3-Embedding-8B & 1.000 & 1.000 & 1.000 \\
& SFR-Embedding-2\_R & 1.000 & 1.000 & 1.000 \\
\midrule
POJ
& C2LLM-7B & \textbf{0.999} & 0.989 & 0.978 \\
& jina-code-embeddings-1.5b & 0.998 & 0.984 & 0.969 \\
& Linq-Embed-Mistral & 0.996 & \underline{0.971} & 0.943 \\
& Octen-Embedding-8B & \textbf{0.999} & 0.990 & 0.981 \\
& Qodo-Embed-1-7B & \underline{0.995} & \underline{0.971} & \underline{0.941} \\
& Qwen3-Embedding-8B & \textbf{0.999} & \textbf{0.992} & \textbf{0.984} \\
& SFR-Embedding-2\_R & \underline{0.995} & \underline{0.971} & 0.942 \\
\bottomrule
\end{tabular}
\end{table}

\subsection{Manual Investigation}
\changed{The embedding analysis indicates potential WT3/T4 mislabelling but cannot estimate its prevalence or characterise its causes.}
\changed{We therefore manually assess a stratified probability sample to estimate this prevalence under our same-main-function definition and examine how method labelling and automatic pairing contribute to the mismatch.}
\changed{The focus of our investigation is the Weakly Type-3/Type-4 (WT3/T4) population because it has the weakest textual constraint, comprises 95.3\% of the reported true clone pairs, and is commonly used as ground truth for semantic clone detection.}
The pairs labelled as Type-1 or Type-2 clone pairs are unproblematic as they are identical or only differ in their whitespace or identifiers and therefore are likely true clone pairs.

\subsubsection{Sampling Strategy}
\changed{We sampled only \texttt{tagged-tagged} pairs classified as WT3/T4, so the 406 selected pairs form a probability sample of the 8,381,611 eligible pairs in the downloaded release, not of the two exemplar-containing pair categories.}
\changed{The sample-size calculation used the reported WT3/T4 population $N=8{,}498{,}894$, a conservative expected proportion $p=0.5$, $z=1.96$, and a five-percentage-point margin of error $e=0.05$.}
\changed{The finite-population formula $n=N z^2p(1-p)/(e^2(N-1)+z^2p(1-p))$ gives 384.143 and therefore a minimum of 385 pairs after upward rounding. This is a precision calculation for a proportion, not a power analysis.}
\changed{For each functionality $h$, we independently rounded the proportional allocation $X_h=385N_h/8{,}498{,}894$ to the nearest integer, using the reported functionality populations $N_h$. The rounded allocations sum to 386.}
\changed{Within each functionality, eligible pairs were randomly ordered and the first $X_h$ pairs were selected.}
\changed{Because $X_h$ rounded to zero for 20 small strata, the first randomly ordered pair from each was added, producing a disproportionate stratified sample of 406 pairs with every functionality represented.}
\changed{The random seed was not recorded, so the realised pair identifiers are archived with the artefact.}
\changed{The population analysis uses all 406 pairs and weights each functionality by its share of the actual 8,381,611-pair WT3/T4 sampling population.}

\subsubsection{Manual Checking}
\label{sec:manual-checking}
\changed{We checked whether each of the 406 pairs in the stratified sample is indeed a true clone pair where the methods are functionally similar.}
We use the following typical definition of being functionally similar:
\begin{quote}
\emph{Two code snippets are considered functionally similar when they achieve the same result or perform the same task, even if they differ in syntax, structure, or the specific steps they take to accomplish that task.}
\end{quote}
\changed{We operationalised this definition by requiring the shared result or task to represent the main overall purpose of both methods.}
\changed{Two methods that merely share a subordinate functionality, but have different main purposes, were therefore not considered functionally similar.}
\changed{The reported mislabelling rate is conditional on this operational definition. A more permissive definition based on any shared functionality would yield a different rate.}
\changed{Both authors applied the two questions in the protocol: they first judged how each method related to the named functionality and then judged whether the pair performed the same task.}
\changed{Where an initial judgement exposed an unclear boundary, the authors clarified during their consensus discussion whether the difference altered the performed task or only its implementation. These discussions applied the existing definition rather than introducing an additional criterion.}
As the disproportionate stratified sample is a superset of the stratified sample, we checked the additional 20 pairs at the same time.
We also checked whether the methods in the pair have been labelled correctly if we apply a stricter criterion: 
\textit{Does the method implement the functionality \textbf{as its main or only purpose}?}
For example, when investigating the ``Copy File'' functionality, methods that (A)~encode, convert or rewrite, (B)~read or write to a stream that is not a file, or (C)~are (unit) tests, are not considered fulfilling the requirement.
\changed{Each author independently examined the same two complete method snippets and functionality specification and recorded a pair label together with separate left/right method labels before any discussion.}
\changed{The method labels distinguish \emph{major}, \emph{incomplete}, \emph{minor}, \emph{wrong}, and \emph{test}. A \emph{major}/\emph{major} classification is usual for a true pair but does not determine the pair label mechanically.}
\changed{For example, in one ``Secure Hash'' disagreement, both methods computed an MD5 digest but differed in UTF-8 handling, hexadecimal padding, logging, and exception handling. The discussion treated these as implementation differences and retained the pair as functionally similar.}
\changed{For the 42 pair-label disagreements, the authors discussed the concrete snippets, clarified how the common definition applied to each boundary case, and recorded one consensus label. The artefact preserves the original independent labels and consensus decisions.}

\begin{table}
\caption{Results of the manual validation of the 386/406 clone pairs after disagreement resolution.
Each row shows the functionality, the number of pairs in the sample (Pairs \#), the number of disagreements between the human investigators (Human \# Dis.), the number of confirmed true positives, and the percentage of true positives (\%) for that functionality.
Additions from the disproportionate stratified sampling are marked with $^*$.
Also shown are the number of disagreements with GPT-4o (LLM \# Dis.).}
\label{tab:results406}
\centering
\begin{tabular}{l|r|r|rr|r}
\toprule
\textbf{Functionality} & \textbf{Pairs} & \textbf{Human} & \multicolumn{2}{c|}{\textbf{True Positives}} & \textbf{LLM} \\
 & \multicolumn{1}{c|}{\textbf{\#}} & \multicolumn{1}{c|}{\textbf{\# Dis.}} & \multicolumn{1}{c}{\textbf{~~~\#}} & \multicolumn{1}{c|}{\textbf{\%}} & \multicolumn{1}{c}{\textbf{\# Dis.}} \\
\midrule
Download From Web              &  18 &  3 &  0 & 0\% & \\
Secure Hash                    &  37 & 22 &  9 & 24\% & 7 \\
Copy File                      & 211 &  1 & 10 & 5\% & 3 \\
$^*$Decompress zip archive     &   1 &    &  1 & 100\% & \\
Connect to FTP Server          &   1 &    &  0 & 0\% & \\
$^*$Bubble Sort Array          &   1 &    &  0 & 0\% & \\
$^*$Setup SGV                  &   1 &    &  0 & 0\% & \\
$^*$Setup SGV Event Handler    &   1 &    &  0 & 0\% & \\
Execute update and rollback    &  12 &  3 &  0 & 0\% & \\
$^*$Initialize Java Eclipse Project & 1 & 1 & 0 & 0\% & \\
$^*$Get Prime Factors          &   1 &    &  0 & 0\% & \\
$^*$Shuffle Array in Place     &   1 &    &  1 & 100\% & \\
Binary Search                  &   4 &    &  2 & 50\% & 1 \\
$^*$Load Custom Font           &   1 &    &  0 & 0\% & \\
$^*$Create Encryption Key Files &  1 &    &  0 & 0\% & \\
$^*$Play Sound                 &   1 &    &  0 & 0\% & \\
$^*$Take Screenshot to File    &   1 &    &  0 & 0\% & \\
$^*$Fibonacci                  &   1 &  1 &  1 & 100\% & \\
$^*$XMPP Send Message          &   1 &    &  0 & 0\% & \\
$^*$Encrypt To File            &   1 &    &  0 & 0\% & \\
Resize Array                   &   4 &    &  0 & 0\% & \\
Open URL in System Browser     &   3 &    &  0 & 0\% & \\
$^*$Open File in Desktop Application &  1 &    &  0 & 0\% & \\
$^*$GCD                        &   1 &    &  1 & 100\% & \\
Call Method Using Reflection   &   4 &    &  0 & 0\% & \\
Parse XML to DOM               &   1 &    &  0 & 0\% & \\
$^*$Convert Date String Format &   1 &    &  0 & 0\% & \\
Zip Files                      &  44 &  6 &  0 & 0\% & \\
File Dialog                    &   5 &    &  0 & 0\% & \\
Send E-Mail                    &   1 &    &  0 & 0\% & \\
CRC32 File Checksum            &   2 &    &  0 & 0\% & \\
Execute External Process       &   5 &    &  0 & 0\% & \\
Instantiate Using Reflection   &  17 &  3 &  0 & 0\% & \\
Connect to Database            &   1 &    &  0 & 0\% & \\
Load File into Byte Array      &   1 &    &  0 & 0\% & \\
$^*$Get MAC Address String.    &   1 &  1 & 0 & 0\% & \\
Delete Folder and Contents.    &   1 &    &  1 & 100\% & \\
Parse CSV File                 &   1 &    &  0 & 0\% & \\
Transpose a Matrix             &   6 &    &  0 & 0\% & \\
Extract Matches Using Regex    &   6 &    &  0 & 0\% & \\
$^*$Copy Directory             &   1 &  1 &  0 & 0\% & \\
$^*$Test Palindrome            &   1 &    &  1 & 100\% & 1 \\
Write PDF File                 &   1 &    &  0 & 0\% & \\
\midrule
Total & 406 & 42 &27 & 6.7\% & \\
\bottomrule
\end{tabular}
\end{table}

\subsubsection{Results}
Table~\ref{tab:results406} shows the results of the manual validation after disagreement resolution for each of the 43 functionalities.
For each functionality, the number of disagreements between the two human investigators, the number of true positives, and the percentage of true positives for that functionality after the disagreement resolution are shown.
The table contains the stratified random sample of 386 pairs, plus the 20 additional pairs from the disproportionate stratified sampling (marked with $^*$).
In the following, we discuss the stratified sample of 386 pairs and the full sample of 406 pairs, including the additional 20 pairs separately.

\changed{From the 386 allocated pairs\footnote{The samples and the results of the manual validation are available at \url{https://github.com/jkrinke/BigCloneBench-Sample-Validation}.}, 19 were labelled as true positives by both authors, 329 were labelled as false positives by both authors, and 38 were labelled differently by the authors.}
\changed{The observed inter-rater agreement is high (90.2\%), while the expected agreement under Cohen's Kappa is 81.8\%.}
\changed{Because there is a high imbalance of the labels (329 out of 386 pairs were labelled as false positives by both judges), we use Gwet's AC1 statistic~\citep{Gwet2008AC1}, which is less affected by imbalanced data and marginal probability than Cohen's Kappa~\citep{Wongpakaran2013}.}
\changed{Using Gwet's AC1, we obtain a value of 0.880, indicating an `almost perfect' ($>$0.8) agreement between both raters.}
\changed{The agreement coefficient quantifies the repeatability of the labels under the operational rule. It does not establish that the chosen construct is objectively correct.}
Note that human investigators have different interpretations of functional similarity and the decisions are highly subjective~\citep{kapser2007subjectivity,charpentier2017raters,Farmahinifarahani2019,Middleton2022}.
Cohen's Kappa and Gwet's AC1 have a different concept of agreement by chance and therefore implications for interpreting agreement scores.
\changed{Here, Cohen's Kappa score is only 0.460, which indicates `moderate' agreement between both raters.}
This low Kappa score is caused by the high imbalance of the labels, a phenomenon known as the Kappa paradox~\citep{Feinstein1990}, and we will therefore only use Gwet's AC1 in the following.

\changed{From the complete sample of 406 pairs, 23 (19+4) were labelled as true positives by both authors, 42 (38+4) were labelled differently by the authors, and 341 (329+12) were labelled as false positives by both authors.}
\changed{The observed inter-rater agreement is 89.7\%, while the expected agreement under Cohen's Kappa is 80.2\%.} Gwet's AC1 is 0.872, again indicating an `almost perfect' agreement between both raters.

Both raters have discussed each of the 42 pairs that they labelled differently and have come to an agreement that only 4 of them are true positives.
\changed{For the 386 allocated pairs alone, disagreement resolution produced 22 true positives and 364 false positives, an unweighted mislabelling rate of 94.3\%.}
\changed{The final human consensus for the 406 pairs contains 27 true positives and 379 false positives, an unweighted mislabelling rate of 93.3\%.}
\changed{Weighting each functionality by its share of the eligible WT3/T4 sampling population gives an estimated mislabelling rate of 93.9\%.}
\changed{For strata containing at least two sampled pairs, we estimate the variance using the usual stratified simple-random-sampling estimator with a finite-population correction.}
\changed{Because a within-stratum variance cannot be estimated from one observation, we use the maximum possible variance of a binary outcome for every singleton stratum, giving a conservative 95\% confidence interval of 91.5\%--96.2\%.}
\changed{This interval treats the adjudicated labels as fixed and therefore quantifies sampling uncertainty, but not the uncertainty arising from the interpretation of functional similarity.}

\changed{The most disagreements between the two human investigators occurred for the functionality ``Secure Hash'' (22 disagreements out of 37 pairs in the sample).}

\changed{The high number of identified false positives indicates that the construction of the WT3/T4 ground truth yields substantial mislabelling under our operational definition of functional similarity.}
\changed{This mislabelling threatens the validity of evaluations that treat WT3/T4 as ground truth for same-main-function semantic similarity.}
\changed{To assess the sensitivity of this estimate to disagreement resolution, we also use only the 341 pairs that both human investigators independently labelled as false positives before discussion and conservatively treat all 42 disagreements as true positives (i.e., assuming that the 42 disagreement pairs are really clone pairs).}
\changed{Applying the same population weights and conservative variance calculation yields an estimated mislabelling rate of 84.4\% (95\% confidence interval: 81.4\%--87.4\%).}
\changed{Thus, the conclusion that a large majority of the sampled pairs are mislabelled does not depend on the adjudication of disagreements, although their adjudication affects the estimated magnitude.}

\begin{quote}
	\changed{\textbf{Under our operational definition of functional similarity, the population-weighted stratified sample estimates that 93.9\% of the eligible WT3/T4 sampling population is mislabelled (conservative 95\% confidence interval: 91.5\%--96.2\%).}}
\end{quote}

When looking at the three functionalities with the most samples, the results are varied.
For the functionality with the most samples (``Copy File''), only 4.7\% of the samples are confirmed true positives.
Interestingly, for the functionality with the second most samples (``Zip Files''), none of the samples are confirmed true positives.
For the functionality with the third most samples (``Secure Hash''), 24.3\% of the samples are confirmed true positives.
\changed{These differences are descriptive and do not show that greater implementation complexity made the ``Zip Files'' pairs more likely to be true positives.}
\changed{We therefore do not infer a cause for the differences among the three functionalities.}

\changed{The 406 sampled pairs involve 779 unique methods, of which 32 appear in more than one pair.}
\changed{Of these methods, 189 were consistently labelled as true positives by both authors, 454 as false positives by both, and 135 differently by the two authors.}
\changed{For one further method, one author assigned a false-positive label in one appearance and a true-positive label in another.}
\changed{After excluding this method, the observed inter-rater agreement is 82.6\%, the expected agreement under Cohen's Kappa is 54.7\%, and Gwet's AC1 is 0.689, which indicates `substantial' agreement between both raters.}

Again, the number of false positives is very high, which casts doubts on the ground truth quality created from the manual labelling of methods.
However, note that the manual validation asked for a stricter criterion than the original ground truth construction.
\changed{Applying the observed ratio of 189/779 to all 15,290 methods labelled as true positives for a specified functionality would give approximately 3,710 correctly labelled methods (24.3\%).}
\changed{This extrapolation is illustrative only, because the methods were obtained from a stratified sample of pairs, not an equal-probability sample of methods.}
\changed{If all 779 methods were paired with one another and a pair were treated as correctly labelled if and only if both methods were among the 189 correctly labelled methods, then $(189*188/2)/(779*778/2) = 5.9\%$ of pairs would be correctly labelled, corresponding to 94.1\% mislabelling.}
\changed{This illustrative value is similar to the population-weighted 93.9\% pair-level result, but ignores the benchmark's within-functionality pairing and is not an independent estimate of the actual mislabelling rate.}
The quadratic pairing of methods turns a bad labelling of methods into a very bad labelling of clone pairs, as the number of pairs grows quadratically with the number of methods.

When looking at the manual classification of each of the 779 methods, the first reviewer classified 285 as \emph{wrong}, i.e., the method does not implement the specified functionality, 47 as \emph{incomplete}, i.e., not the full specified functionality has been implemented, 222 as \emph{minor}, i.e., the specified functionality is not the primary functionality, and 25 as \emph{tests}, i.e., the specified functionality is used as a test. Only 200 out of 779 have been classified as implementing the specified functionality as the primary (\emph{major}) functionality and therefore as a confirmed true positive.
\changed{The second reviewer consistently classified 302 methods as wrong, 14 as incomplete, 134 as minor, 13 as test, and 313 as major and therefore as confirmed true positives.}
\changed{For the remaining three methods, repeated appearances received major and wrong, minor and wrong, and incomplete and minor classifications, respectively.}
\changed{Five pairs received a final human false-positive label even though both reviewers independently classified both methods as implementing the specified functionality as their major functionality.}
\changed{For three of these pairs, both reviewers also independently classified the pair as a false positive.}
\changed{The remaining two received that pair label after an initial disagreement was resolved.}
\changed{For example, in one ``Zip Files'' pair, one method archives the non-directory files of a supplied directory, whereas the other constructs a launch-profile package containing one designated file under a fixed archive entry with explicit size and checksum metadata.}
\changed{Both use ZIP creation as a major functionality, but their overall tasks differ: one performs general directory archiving, while the other constructs a particular application package.}
\changed{The pair was therefore classified as a false positive under our operational definition.}

Because the classification of the functionality of a method is highly subjective, the disagreement on the classification of methods is much higher.
It is challenging to decide if a method's implementation of a specified functionality is complete or if the functionality is only a minor one.
As seen with the second reviewer, this occurs even when the same investigator has to classify the same method more than once.

Surprisingly, for 219 methods, both reviewers classified the method as wrong, i.e., that the method does not have the specified functionality at all.
This occurred for methods with the functionalities `Download From Web' (6 methods), `Copy File' (200), `Resize Array' (7), `Zip Files' (4), and `Transpose a Matrix' (2).
For example, the clearest mislabelling is the two methods for `Transpose a Matrix':
One method actually computes the distances between a number of points, and the other generates a random undirected graph.
For both methods, the data reveals that both judges must have labelled the two methods as true for the functionality `Transpose a Matrix', even though both methods do not transpose a matrix.

The ratio of 219/779 (28\%) of methods labelled as not having the specified functionality raises additional concerns about the labelling process employed during the BigCloneBench construction.
\changed{Under the relaxed criterion that the specified functionality need only be implemented as a minor functionality, an illustrative calculation assumes that all 779 methods are paired with one another and that a pair is mislabelled if and only if at least one method is among the 219 classified as wrong.}
\changed{Under these assumptions, $1-(560*559/2)/(779*778/2) = 48\%$ of clone pairs would be mislabelled.}
\changed{This is not an estimate of the actual pair-level mislabelling rate: BigCloneBench pairs methods within functionalities, and methods wrongly labelled for a specified functionality may still be functionally similar to each other.}
\changed{However, the focus of this paper is the mislabelling of clone pairs and, therefore, a further analysis of the methods for mislabelling is outside the scope of this paper.}

\subsection{Post-hoc Comparison with an LLM}

\changed{The human consensus labels and the resulting population estimate were obtained without an LLM.}
\changed{We then used LLM explanations to identify cases for post-hoc scrutiny among the 406 pairs and measured the effect of any accepted corrections.}
\changed{Because the authors reconsidered disagreements after seeing the model's explanations, this comparison is not an independent validation or a separate source of ground truth.}

\begin{figure}
\footnotesize
\begin{verbatim}
You are an experienced software engineer proficient
in analysing source code.

Your task is to analyse two code snippets and check
if they are functionally similar. Two code snippets
are considered functionally similar when they
achieve the same result or perform the same task,
even if they differ in syntax, structure, or the
specific steps they take to accomplish that task.

Code Snippet 1: """
{contentA}
"""

Code Snippet 2: """
{contentB}
"""

Compare the two Java code snippets. Answer
"YES-SIMILAR" if they are functionally similar.
Answer "NO-NOT-SIMILAR" if they are not functionally
similar. Answer "DONT-KNOW" if it is not clear if
they are similar.

Format:
ANSWER: <answer>
EXPLANATION: <explanation>
\end{verbatim}
\caption{Instructions for the LLM}
\label{fig:llm}
\end{figure}

\changed{We used OpenAI's \texttt{gpt-4o-2024-08-06} model through its API. Exploratory comparisons of models and prompts informed this choice, but their records and selection criterion were not retained, so we do not claim that the selected configuration was optimal.}
\changed{The prompt in \Cref{fig:llm} contains the general definition of functional similarity but does not impose the major/minor method-classification criterion used in the human protocol.}
\changed{Each request provided the two complete Java snippets in a fixed order, but not the BigCloneBench functionality tag, pair label, or human decisions.}
\changed{We set \texttt{temperature=0} and \texttt{top\_p=0} and ran the comparison five times.}
\changed{These settings reduce sampling variation but do not guarantee deterministic output~\citep{Ouyang2025}.}
\changed{The replication artefact contains the five raw logs, exact prompt, recorded configuration and its limitations, checksums, and an offline reproducer. No further API request is needed to reproduce the reported comparison.}

\changed{Four runs produced identical labels for all 406 pairs, while the fifth differed on one pair. Majority voting gave 21 similar and 385 not-similar labels.}

\begin{table}[tbp]
\centering
\caption{\changed{Comparison of the human consensus reached before the LLM analysis with the LLM majority labels.}}
\label{tab:human-llm}
\begin{tabular}{lrrr}
\hline
 & \multicolumn{2}{c}{LLM majority} & \\
Human consensus & Similar & Not similar & Total \\
\hline
Similar     & 18 &   9 &  27 \\
Not similar &  3 & 376 & 379 \\
\hline
Total       & 21 & 385 & 406 \\
\hline
\end{tabular}
\end{table}

\changed{As shown in \Cref{tab:human-llm}, the human consensus and LLM majority agreed on 394 pairs and disagreed on 12, giving 97.0\% observed agreement and Gwet's AC1 of 0.967.}
\changed{The five runs therefore show high repeatability within the recorded configuration, but the high human--LLM agreement does not establish correctness, particularly given the class imbalance.}

\changed{The authors re-examined the 12 disagreements with access to the model's explanations and retained responsibility for every final label.}
\changed{For seven ``Secure Hash'' pairs labelled not similar by the LLM, we retained the human labels because both methods had secure hashing as their main task under our operational definition, notwithstanding differences in hash algorithm or encoding.}
\changed{A narrower task requiring the same algorithm or output format could draw a different boundary, but no such task was evaluated here.}
\changed{For three ``Copy File'' pairs labelled similar by the LLM, we retained the human labels because the methods had materially different main tasks. For example, one comparison paired a general method with a complete file-copying program.}
\changed{We accepted the challenges to the ``Binary Search'' and ``Test Palindrome'' pairs: the former differed in data structures and additional functionality, while the latter contained a logical error overlooked during the initial human assessment.}

\changed{Thus, ten challenges were rejected and two were accepted, increasing the identified false positives from 379 to 381.}
\changed{Applying the same population weights after these changes gives a sensitivity estimate of 94.1\% (conservative 95\% confidence interval: 91.8\%--96.4\%), compared with the primary pre-LLM estimate of 93.9\% (91.5\%--96.2\%).}
\changed{The 0.2-percentage-point change does not materially alter the conclusion, but the two corrected labels show the limited diagnostic value of model explanations for exposing details that merit human re-examination.}

\section{Investigate the Impact of BCB Ground Truth Quality on the Literature\label{sec:literature-impact}}

\changed{Having estimated the scale of mislabelling in the sampled WT3/T4 population and examined the stability of the manual conclusion, we use a literature search to estimate how extensively published research is exposed to the resulting validity threat through the use of WT3/T4 for evaluation or machine learning.}

\changed{Traditional methods for clone detection are often evaluated on the full BigCloneBench, including the WT3/T4 subset.}
Such papers report extremely low recall for the WT3/T4 subset and usually argue that the complexity for such clone pairs is the reason for the low recall and that the presented approach is not targeted at WT3/T4 clones.
\changed{The real recall of these traditional approaches may therefore be higher, as our manual study estimates that most pairs in the sampled WT3/T4 population are not clones under our operational definition.}

\begin{figure}[tbp]
    \centering
    \includegraphics[width=\textwidth]{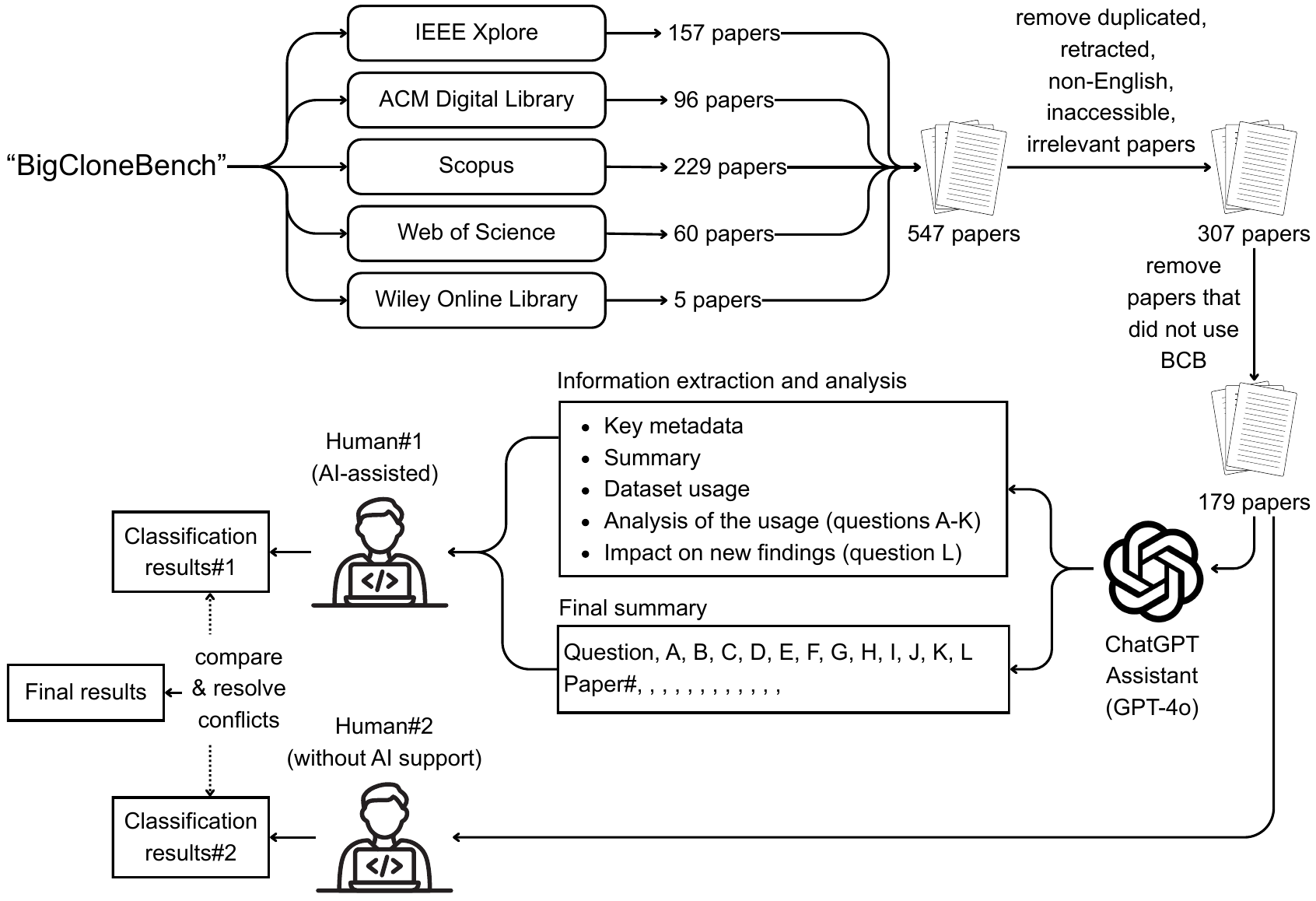} \caption{Process for collecting and analysing papers that have used BigCloneBench.}
    \label{fig:literature_study}
\end{figure}

\changed{\Cref{fig:literature_study} summarises the collection, LLM-assisted screening, and two-human classification described below.}

\subsection{Paper Collection}

\changed{Following the methodology of \citet{Pasuksmit2024}, we searched IEEE Xplore, ACM Digital Library, Scopus, Web of Science, and Wiley Online Library for ``BigCloneBench'' in the full text or metadata, covering all indexed papers up to 20 March 2025 and including adjacent tasks such as clone search~\citep{Ragkhitwetsagul2019}.}
\changed{The search returned 547 records: 157 from IEEE Xplore, 96 from ACM Digital Library, 229 from Scopus, 60 from Web of Science, and 5 from Wiley Online Library.}
\changed{After manually removing 226 duplicates, 2 retracted papers, 4 non-English papers, 3 inaccessible papers, 5 irrelevant records such as contents pages of conference proceedings, messages from chairs, or abstracts, our own paper~\citep{Krinke2022}, and the original BigCloneBench paper~\citep{Svajlenko2014}, 305 papers remained.}
\changed{We added two papers that used BigCloneBench but were not indexed by these sources, giving 307 downloaded PDFs.}

\subsection{Paper Analysis}

\changed{We used an LLM to extract information and provide an initial classification of the 307 papers, not as a neutral or independent judge.}
\changed{For each PDF, the first prompt requested factual information with supporting quotations and a provisional assessment of whether the observed WT3/T4 mislabelling affected the paper's results, while the second converted the answers into a CSV-like summary.}
\changed{The 93.3\% value supplied at this stage, rendered as 93.35\% in the verbatim prompt in the appendix, is the unweighted human-consensus proportion of 379/406 available to the literature analysis, not the population-weighted estimate reported in \Cref{sec:manual-checking}.}
\changed{We refined the prompts by repeatedly running them on a random sample of 20 papers and manually inspecting the responses, and the final prompts are reproduced in the appendix.}
\changed{During this development, we tried several LLMs and selected GPT-4o because the alternatives either lacked direct PDF-analysis support or produced lower-quality responses.}
\changed{The responses for all papers were saved and manually checked by the first author.}
\changed{The request script used the OpenAI Assistants API with the \allowbreak{}\texttt{gpt-4o-2024-08-06} model.}
\changed{The script also records the assistant instructions, PDF file-search input, two sequential prompts, one run per prompt, and \allowbreak{}\texttt{temperature} and \allowbreak{}\texttt{top\_p} values of $10^{-14}$. It does not specify a seed or output-token limit, the access dates were not recorded, and unsuccessful runs terminate rather than being retried.}

\changed{The LLM-assisted screening excluded 19 surveys and 109 papers that did not use BigCloneBench as a dataset, leaving 179 papers for classification according to whether WT3/T4 mislabelling could weaken claims based on their BigCloneBench results.}
\changed{Human \#1 checked every paper with the LLM's suggested labels and evidence, while Human \#2 classified the papers without LLM assistance.}
\changed{The two investigators discussed the 23 Yes/No disagreements and the one case marked ``?'' by Human \#1 and reached the final consensus reported below.}

\begin{table}[tbp]
\caption{Inter-rater reliability, Cohen's Kappa, and Gwet's AC1 across AI and human evaluators on paper classification}
\label{tab:rater_agreement}
\centering
\begin{tabular}{lrrr}
\toprule
\textbf{Pair (Rater 1 vs Rater 2)} & \textbf{Agreement} & \textbf{Cohen's Kappa} & \textbf{Gwet's AC1} \\
\midrule
AI vs. Human 1 (Assisted) & 77.5\% & 0.085 & 0.710 \\
AI vs. Human 2 (Unassisted) & 71.5\% & 0.089 & 0.607 \\
\textbf{Human 1 vs. Human 2} & \textbf{87.1\%} & \textbf{0.672} & \textbf{0.788} \\
\midrule
AI vs. Final Consensus & 73.2\% & 0.097 & 0.636 \\
Human 1 vs. Final Consensus & 91.0\% & 0.766 & 0.854 \\
Human 2 vs. Final Consensus & 96.1\% & 0.906 & 0.933 \\
\bottomrule
\end{tabular}
\end{table}

Table~\ref{tab:rater_agreement} presents the agreement metrics across the different pairs.
\changed{Comparisons involving Human \#1 exclude the one classification recorded as ``?'' and use 178 papers, whereas the other comparisons use all 179 papers.}
The base classification rates (percentage of papers classified as affected by the issue) were drastically different among raters: the AI classified 174 (97.2\%) of the papers as affected, Human \#1 classified 137 (76.5\%), Human \#2 classified 125 (69.8\%), and the final consensus settled at 128 out of 179 papers (71.5\%). 

\changed{These figures describe the classification process, but the design does not isolate the effect of LLM assistance from differences between the two human reviewers.}
\changed{The assisted reviewer agreed more closely with the LLM than the unassisted reviewer (77.5\% versus 71.5\%), whereas the unassisted reviewer agreed more closely with the final consensus (96.1\% versus 91.0\%).}
\changed{In eleven cases involving low reported WT3/T4 recall, the LLM and assisted reviewer initially classified the paper as affected, while the unassisted reviewer argued that low recall meant the model had not successfully learned the flawed ground truth.}
\changed{The consensus accepted the latter interpretation and did not classify these eleven papers as potentially affected.}
\changed{This was a judgement about the papers' claims, not a validation of their WT3/T4 measurements.}
\changed{Low recall does not by itself rule out bias in those measurements or their interpretation.}
\changed{This pattern is compatible with anchoring to the LLM suggestions, but cannot establish a causal automation-bias effect.}

\changed{The LLM's highly skewed 97.2\% positive classification rate also affects the interpretation of the agreement measures.}
\changed{Although its observed agreement with the human reviewers was 71.5\%--77.5\%, Cohen's Kappa was only 0.085--0.089 because it estimates chance agreement from the raters' marginal classification rates, illustrating the Kappa paradox under class imbalance~\citep{Feinstein1990}.}
\changed{Gwet's AC1 uses a different model of chance agreement and is less affected by this imbalance, resulting in values of 0.607--0.710~\citep{Gwet2008AC1,Wongpakaran2013}.}
\changed{The two coefficients therefore describe agreement differently, and neither establishes the correctness of the LLM classifications.}

\bigskip
\changed{The manual classification consensus identified 128 of the 179 papers (71.5\%) whose claims based on BigCloneBench's WT3/T4 pairs could potentially be weakened by mislabelling.}
\changed{This count is not an exhaustive measure of exposure to mislabelled WT3/T4 data, and a negative classification does not establish that a paper's measurements are unaffected.}
\changed{The corpus includes cross-dataset studies. For example, \citet{Arshad2022} evaluated CodeBERT on a BigCloneBench subset, SemanticCloneBench, and an Android dataset.}
\changed{This count concerns potential threats to claims based on BigCloneBench results under the classification described above.}
\changed{Evidence from other datasets may reduce the dependence of a paper's broader conclusions on BigCloneBench, but does not remove the validity threat to measurements obtained from its WT3/T4 ground truth.}

\changed{These papers often report high recall or F1 scores.}
For example, an often cited paper introducing ASTNN~\citep{Zhang2019a} reports a precision of 99.8\%, a recall of 88.3\%, and an F1 score of 93.7\% for the WT3/T4 subset (and 99.8\%, 88.4\%, and 93.8\% for the full dataset), as measured on balanced subsets consisting of 40,000 pairs.
\changed{To illustrate the sensitivity of these evaluations rather than estimate clean-data performance, we hold the reported predictions fixed and apply our observed 93.3\% mislabelling ratio as a pessimistic scenario.}
\changed{This gives an adjusted precision of 6.7\% ($0.998 \cdot 0.067$) and, with recall held fixed, an adjusted F1 score of 12.4\%.}
\changed{The calculation cannot estimate performance after retraining on corrected labels, which could change the learned model and improve performance. Corrected evaluation labels would also change the metrics, so the net effect cannot be inferred from the reported results.}
\changed{In particular, we did not conduct a cross-dataset generalisation experiment and therefore cannot conclude that these approaches overfitted their training data.}

\begin{quote}
\textbf{Any machine learning approach that relies on the WT3/T4 subset of BigCloneBench's ground truth for training and evaluation risks modelling dataset artefacts rather than semantic similarity. Our literature analysis resulted in an estimate of 128 papers whose results could be threatened in their validity. It does not definitively invalidate every individual study, but it highlights a systemic risk in recent semantic clone detection research.}
\end{quote}

\changed{Because the investigators and LLM produced different classifications, the final consensus count remains subject to classification uncertainty.}
\changed{In a hypothetical scenario where up to 20\% of the 128 papers classified as potentially affected were false positives, more than 100 papers would remain potentially affected.}
\changed{Our analysis does not and cannot determine whether every one of these papers is invalid.}
\changed{It concerns aggregate exposure to a validity threat rather than judgements about individual papers or authors.}
\changed{The need to re-evaluate a substantial body of published work is itself a harm to the field because it weakens the reliability and comparability of the current evidence base and makes genuine advances harder to identify.}
\changed{Until those results are re-evaluated, confidence in current estimates of semantic-clone detection performance is necessarily reduced.}

We also observed during the literature analysis that the majority of the papers do not report on any kind of manual investigation.
There are some papers that include a manual investigation of the results of the approach with the purpose of estimating the approach's precision (as BigCloneBench has been created as a benchmark for recall), but none of them use the results of the manual investigation to validate the ground truth.
Other papers~\citep{Kim2018,Zeng2019,Zhang2021} investigate only a small sample of results and compare them to the ground truth.
It should be noted that we encountered no paper using the WT3/T4 ground truth of BigCloneBench under the assumption that pairs only need to share a minor functionality to be labelled as true pairs.

\begin{quote}
\textbf{We ask the community not to focus on the threatened papers, but to look forward and not to repeat mistakes in future work by critically evaluating the quality of the datasets and benchmarks used for evaluation and training.}
\end{quote}

\section{Further Observations and Ecosystem Impact\label{sec:observations}}

\changed{We next consider whether the measured WT3/T4 mismatch is an isolated labelling defect or part of a wider set of limitations, deriving eight further observations from the published data before synthesising their implications in the Discussion.}

\subsection{Quality of the Method Labelling}

It has been reported~\citep{Svajlenko2017,Svajlenko2021} that at least one judge disagreed with the others for 14.5\% of the 9,533 methods that have been labelled by more than one judge.
Moreover, 68,400 methods (88\% of all methods) have been labelled by a single judge only, which affects the quality of the labelling.
\changed{Agreement among multiple judges can be used to assess annotation reliability, which is a prerequisite for, but does not itself establish, validity~\citep{Artstein2008}.}
\citet{Svajlenko2017} estimated that 14.5\% of the methods tagged by only a single judge are affected by subjectivity or tagging errors, and \citet{Svajlenko2021} extrapolated the average disagreement to estimate that at least 15\% of the clones across the benchmark are subjective or have validation errors.
However, manual validation of clone pairs is also subjective~\citep{kapser2007subjectivity,charpentier2017raters,Farmahinifarahani2019,Middleton2022}, and the existence of a dissenting judge does not imply that the final label is wrong.

\begin{quote}
	\textbf{\makebox[18ex][r]{(Observation 1)~}88\% of the methods in BigCloneBench have been labelled by a single judge, and for methods with multiple judges, 14.5\% of the methods have been labelled differently by the judges.}
\end{quote} 

\subsection{Incomplete Ground Truth}

Most importantly, the ground truth does not make any assumption about method pairs where the methods appear in sets for different functionalities.
\changed{For example, for a method pair $(m, n)$ with $(m \in X_i \cup P_i) \land (n \in X_j \cup P_j) \land (i \neq j)$, it \emph{cannot} be assumed that $(m, n)$ is a non-clone pair.}
This is highlighted by the fact that BigCloneBench allows methods to be in the true positive sets of different functionalities and indeed has 209 of such methods $m$ with $(m \in P_i) \land (m \in P_j) \land (i \neq j)$.

\begin{quote}
	\textbf{\makebox[18ex][r]{(Observation 2)~}BigCloneBench does not contain information about pairs for different functionalities.}
\end{quote} 

\changed{Any assumption that $(m, n)$ is a non-clone pair solely because the methods are from two different functionalities is not valid.}
Consider the two functionalities for ``Copy File'' and ``Copy Directory''.
The functionality of copying a file is part of the functionality ``Copy Directory'' which copies a directory and its contents.
At least 77 methods in the set of true positives for the functionality ``Copy Directory'' invoke a \texttt{copyFile} method.
Therefore, all methods in the set of true positives for the functionality ``Copy Directory'' could be considered clones of methods for the ``Copy File'' functionality.
Indeed, 19 methods in the set of true positives for the functionality ``Copy Directory'' have also been judged to be true positives for the ``Copy File'' functionality, and only three methods in the set of true positives for the functionality ``Copy Directory'' have been judged to be false positives for the ``Copy File'' functionality.
However, it appears that the heuristic for the functionality ``Copy File'' has not retrieved most of the methods that have been retrieved for the functionality ``Copy Directory''. 

\begin{quote}
	\textbf{\makebox[18ex][r]{(Observation 3)~}BigCloneBench does contain true but unlabelled clone pairs for different functionalities.}
\end{quote}

The above observation is important because there are datasets that are created with the assumption that methods for different functionalities are not clones of each other, and for which the assumption holds.
Examples for such datasets are POJ/OJClone~\citep{Mou2016} or past submissions to the Google Code Jam\footnote{\url{https://codingcompetitions.withgoogle.com/codejam}}.
In such datasets, participants create small programs to solve a given task, which are then checked with defined sets of tests.
It is usually assumed that the solutions of a given task are similar to each other and that the solutions to different tasks are different.

Some of the papers analysed in the literature review follow a similar assumption and, instead of only using the provided ground truth, the approaches construct their own ground truth by considering all methods for different functionalities as false clone pairs.
For example, \citet{Yu2019} use the earlier version of BigCloneBench with 10 functionalities.
They only use the 9,134 methods that have been labelled and split them into 8,134 methods for training, 500 for testing, and 500 for validation.
Each of the sets is then used to create the clone pairs, assuming that methods in the same functionality are true pairs and methods from different functionalities are false pairs.
Their dataset is available\footnote{\url{https://github.com/yh1105/datasetforTBCCD}} and allowed us to investigate the methods and true clone pairs and the false clone pairs used in the machine learning.
The investigation confirmed that the created dataset expanded the ground truth provided by BigCloneBench without considering the issues we observed.
For example, we could confirm the presence of wrongly generated false clone pairs caused by assuming  two methods labelled as true positive for two different functionalities are not clones of each other.
Consider the method no.~22442270, which has been labelled by the human judges as a true positive for the ``Download From Web'' functionality (and as a false positive for the ``Copy File'' functionality).
In Yu et al.'s dataset, this method often appears as part of false clone pairs in the expanded ground truth.
It appears in false clone pairs together with other methods from all other functionalities.
In addition, the dataset is also very imbalanced, with only 14\% of the pairs labelled as true clone pairs (336,498 out of 2,416,586 total pairs).

\citet{Wang2020} seem to follow a similar approach.
\changed{The paper reports that their ground truth contains 336,498 pairs labelled as clones and 2,080,088 labelled as non-clones, although the original ground truth contains 6,164,953 true clone pairs and only 258,574 false clone pairs~\citep{Svajlenko2014}.}
They do not discuss how they have constructed their changed version of the ground truth\footnote{\url{https://github.com/jacobwwh/graphmatch_clone}}.
Interestingly, there are a few papers~\citep{Ji2021,Zhang2023,Chunyan2023,Yu2023} that report the same number of 2,080,088 false clone pairs, but they also report the original contradicting number of around 260,000 (or precisely 279,032) false clone pairs in descriptions of BigCloneBench.
By downloading the datasets for two papers~\citep{Wang2020,Yu2023}, we were able to confirm that they used the exact same data for training, testing, and validation.

The above discussion shows that machine learning approaches using BigCloneBench are at risk when attempting to create complete ground truths, in particular when attempting to label method pairs with methods from different functionalities.

Of lesser importance is the observation that the ground truth is even incomplete within the same functionality.
As discussed above, the ground truth does not make any assumption on method pairs where both methods have been labelled as false positives for the given functionality.
The reason is that such methods are not clones of any exemplar function of the given functionality, but they could be clones of each other.

\begin{quote}
	\textbf{\makebox[18ex][r]{(Observation 4)~}BigCloneBench does contain unlabelled true and false clone pairs for the same functionalities.}
\end{quote}

\subsection{Class Imbalance and Functionality-Distribution Skew}

\changed{The ground truth exhibits both an extreme imbalance between labelled true and false clone pairs and a highly skewed distribution of labelled methods and pair assignments across the 43 functionalities.}
\changed{For example, 35,948 of the 75,673 labelled method--functionality records in the released dataset concern ``Copy File'' (47.5\%).}
\changed{Of the 14,891 records labelled as true positives, 3,049 concern ``Copy File'' (20.5\%).}
\changed{In the downloaded release, 4,664,949 of the 8,584,153 true clone pairs concern ``Copy File'' (54\%).}
\changed{The six exemplar functions and 3,049 true-positive methods yield 4,664,985 possible unordered pairs.}
\changed{Of these, 36 are absent from the released table, all involving methods with overlapping line ranges in the same source file.}
Over 90\% of all true clone pairs are just for eight functionalities, and 22 out of the 43 functionalities taken together only amount to less than 1\% of all true clone pairs.
\changed{The imbalance is even worse for the labelled false clone pairs, of which 70\% concern the functionality ``Copy File''.} Over 98\% are pairs just for eight functionalities. 32 out of the 43 functionalities taken together only amount to less than 1\% of all false clone pairs.

\begin{quote}
	\changed{\textbf{\makebox[18ex][r]{(Observation 5)~}BigCloneBench exhibits extreme class imbalance between labelled true and false clone pairs and a highly skewed distribution of method and pair assignments across its 43 functionalities.}}
\end{quote}

The five above observations only have a limited impact on the evaluation of clone detectors that only use the clone pairs of the ground truth.
\changed{However, this class imbalance and functionality-distribution skew can affect machine learning approaches for code similarity that learn from BigCloneBench's ground truth.}
\changed{For example, an approach that recognised only whether both methods implement some copy-file functionality could achieve a recall of approximately 55\%, because 4,651,096 of the 8,498,894 true WT3/T4 clone pairs concern ``Copy File''.}

This is in addition to the imbalance between true clone pairs and false clone pairs, where, as discussed above, even poor approaches that always predict true achieve a high recall of 100\% and a high precision of 96.9\%, leading to an F1 score of 98.4\%.

\begin{quote}
	\changed{\textbf{\makebox[18ex][r]{(Observation 6)~}BigCloneBench permits an F1 score of 98.4\% from predicting every labelled pair as a clone and a recall of approximately 55\% from recognising clone pairs only for the ``Copy File'' functionality. Neither score demonstrates that an approach has learned functional similarity.}}
\end{quote}

The imbalance of BigCloneBench has been observed previously, e.g., by \citet{Alam2023}, or by \citet{Li2023} who also tried to repair the imbalance.

\subsection{Repair Attempts and Propagation into Derived Datasets}

Being aware of problems with BigCloneBench, \citet{Li2023} have attempted to repair the dataset.
They attempt not only to fix the imbalance by creating a balanced subset, but they also try to remove wrongly labelled pairs by applying a similarity threshold and by removing all methods that have been labelled by a single person.
However, given that clone pairs in the WT3/T4 category already have a low similarity based on a threshold, there is no evidence that the created subset is better than the original BigCloneBench dataset regarding wrongly labelled WT3/T4 pairs.

The same imbalance problem was observed and mitigated by \citet{kitsios2025} to study how models trained on one functionality of BigCloneBench generalise to other functionalities.
\changed{However, the drop in F1 scores they observed could be influenced by WT3/T4 ground-truth mislabelling, as discussed above.}

\begin{quote}
	\changed{\textbf{\makebox[18ex][r]{(Observation 7)~}The reviewed repair attempts have not been shown to correct WT3/T4 ground-truth alignment with same-main-function similarity.}}
\end{quote}

\changed{A study by \citet{Yu2022} groups MT3 and WT3/T4 as semantic clones and shows that pairs in this combined subset usually contain the same identifier names.}
\changed{BigCloneBench assigns these categories using textual overlap thresholds rather than semantic validation: MT3 pairs share 50--70\% of their normalised lines or tokens, whereas WT3/T4 pairs share less than 50\%.}
\changed{The textual overlap of MT3 pairs can provide a separate basis for treating them as syntactic Type-3 clones, independently of whether their methods have the same main functionality. We therefore exclude MT3 from our population estimate and do not generalise the WT3/T4 estimate to MT3.}
\changed{The same permissive method-labelling and automatic pairing mechanism also produces MT3 pairs, so their suitability as ground truth for same-main-function similarity remains an open empirical question that would require a separate probability sample.}
\changed{In the earlier 10-functionality version used by that study, MT3 accounts for 1.01\% and WT3/T4 for 98.23\% of the true clone pairs, so the combined subset is numerically dominated by WT3/T4.}
Their experiment using a Linear-Model to detect semantic clone pairs in BigCloneBench, only based on identifier names, provides a comparable result to the state-of-the-art ML-based techniques such as ASTNN~\citep{Zhang2019a}, TBCCD~\citep{Yu2019}, and FA~\citep{Wang2020}.
They also report that the performance of the three tools, trained on the original BigCloneBench, on a revised BigCloneBench dataset after abstracting the identifier names drops significantly (F1-score decreases 16\%--27\%).
\changed{If the combined MT3 and WT3/T4 ground truth could be trusted, their results would indicate a threat to external validity: models that perform well when trained and evaluated on BigCloneBench may not perform equally well when semantic clone pairs use different identifier names.}
\changed{However, because their results combine MT3 with potentially mislabelled WT3/T4 ground truth, it is unclear whether the results would be the same with corrected labels, and the study does not independently establish the same-main-function label quality of either category.}

There exist datasets that are derived from BigCloneBench or include BigCloneBench as part of a larger dataset.
\citet{Yu2019} created a dataset for training and evaluating their TBCCD approach, but added additional false clone pairs by assuming that methods from different functionalities are not clones of each other (as discussed above).
The same dataset has been used in other papers~\citep{Wang2020,Yu2023}.
This dataset is similar to the dataset created by \citet{Wei2017} for their CDLH approach, which has not been publicly released.

CodeXGLUE~\citep{Lu2021b} is a collection of datasets for machine learning for code understanding and generation tasks.
It includes datasets for code summarisation, code translation, code completion, and clone detection, for which it includes a modified version of BigCloneBench.
\changed{Because its clone-detection component is a modified version of BigCloneBench, the documented ground-truth alignment concern remains unless its labels are independently validated for the intended construct.}
Moreover, the downloaded version of CodeXGLUE\footnote{\url{https://github.com/microsoft/CodeXGLUE/tree/main/Code-Code/Clone-detection-BigCloneBench}} consists of 9,126 methods, 1,170,339 false clone pairs, and 561,521 true clone pairs.
This is again problematic as the original BigCloneBench has only 258,574 false clone pairs, and it is not clear how the additional false clone pairs have been generated. Moreover, there are only 336,090 unique true clone pairs in the dataset.

We again have checked all false clone pairs that involve the sample method no.~22442270 and found that the dataset contains pairs of that sample to methods in all other functionalities of the older BigCloneBench and some more functionalities that only appear in the newer BigCloneBench.
This observation indicates again potential mistakes in the construction of this dataset, similar to what we have observed for other usages of BigCloneBench.

\changed{Papers may use these derivatives without citing BigCloneBench, so the literature review may not capture all exposed work. We did not estimate this additional population.}

\begin{quote}
	\changed{\textbf{\makebox[18ex][r]{(Observation 8)~}The inspected TBCCD derivative adds negative pairs using an assumption unsupported by BigCloneBench's ground truth, while CodeXGLUE contains unexplained additional negative pairs. Both require independent label validation.}}
\end{quote}

\section{Discussion}

\changed{The primary evidence is the population-weighted estimate of 93.9\% WT3/T4 mislabelling and the adjudication-independent sensitivity estimate of 84.4\% in \Cref{sec:manual-checking}.}
\changed{The literature review classified 128 papers as having claims potentially affected by this validity threat (\Cref{sec:literature-impact}), while the descriptive analyses in \Cref{sec:observations} established arithmetic class- and functionality-frequency baselines and traced unsupported assumptions in two inspected derivatives.}
\changed{The arithmetic baselines demonstrate opportunities for shortcuts, not that any individual approach used them.}
\changed{Consequently, performance on BigCloneBench alone is insufficient evidence that a machine learning approach has learned same-main-function similarity.}

\changed{The construction record identifies three decisions relevant to this mismatch.}
The first decision was to give the human judges a broad interpretation of functional similarity, allowing methods to be labelled as similar to a functionality even if the functionality was a simple and minor element.
The second decision was based on the assumption that two methods labelled with the same functionality would be functionally similar.
\changed{The third decision was to rely on individual human judgements, which are subject to labelling error.}
\changed{Together with the automatic construction of clone pairs, these decisions explain how a recall-oriented benchmark acquired a flawed WT3/T4 positive reference set that is not aligned with same-main-function semantic similarity.}

\changed{This mismatch became consequential when BigCloneBench was repurposed for training and evaluating machine learning-based clone detection approaches.}
Work that used BigCloneBench often demonstrated improved performance over previous work by reporting higher scores in precision, recall, or F1.
\changed{Because the WT3/T4 ground truth is misaligned with same-main-function similarity, such scores do not by themselves establish the claimed level of semantic-clone performance.}
\changed{Our study did not test individual architectures and therefore cannot determine whether reported improvements reflect genuine advances, benchmark-specific properties, or both.}
\changed{Likewise, poor performance on this ground truth cannot establish that an approach is ineffective for a better-aligned task.}
\changed{Re-evaluating affected results entails further researcher effort and computation, including electricity use, although these resource and sustainability implications are not quantified here.}

\changed{The appropriate boundary for functional similarity is task-dependent, and without an established practitioner use case neither our same-main-function boundary nor a more permissive one can be treated as universally appropriate.}
\changed{However, a possible task identified only after inspecting the available labels would not by itself provide evidence of practical relevance, which must instead be established from the intended task and its users.}
\changed{Related studies show that offline evaluations can misrepresent the intended problem or user value when they omit necessary program context or target users~\citep{Risse2025,Kovalenko2020}.}
\changed{A functional-containment task that retrieves methods containing a specified operation is conceivable, but this is not BigCloneBench's stated Type-4 construct of code fragments implementing the same functionality.}
\changed{The permissiveness instead enters its construction, where individual methods may be labelled as containing a target functionality despite additional unrelated tasks and methods with the same label are then paired automatically.}
\changed{The method labels might provide candidates for a separately designed and validated containment benchmark, but this possibility does not validate the existing WT3/T4 pair labels against either BigCloneBench's stated Type-4 definition or the same-main-function construct evaluated here.}

\begin{quote}
\changed{\textbf{We recommend against using BigCloneBench's WT3/T4 subset as ground truth for training or evaluating approaches intended to detect same-main-function semantic similarity.
This recommendation also applies to derived datasets unless their labels have been independently validated for that construct.}}
\end{quote}

\changed{This recommendation is limited to same-main-function use: our manual estimate does not assess traditional recall uses or labels governed solely by textual clone-category thresholds.}
\changed{Repairing the WT3/T4 ground truth would require relabelling methods under a stricter definition and reconstructing the affected pairs. We did not assess the feasibility of such a repair.}
\changed{For same-main-function training and evaluation, selecting and validating an alternative may therefore be more practical (\Cref{sec:datasets}).}

\changed{The scope of this critique is important:}
\begin{quote}
\textbf{\changed{Repurposing BigCloneBench's WT3/T4 subset as ground truth for same-main-function similarity without validating its labels for that task is a misuse of the dataset.}
\changed{Its widespread adoption has harmed the evidence base of semantic clone detection by leaving a substantial body of results in need of re-evaluation.}
\changed{This critique concerns the community's responsibility to check benchmark--task alignment. It is not directed at the creators of BigCloneBench or the authors of individual studies.}}
\end{quote}

\changed{Researchers may inherit assumptions from prior studies without re-ex\-amining dataset construction, limitations, and task suitability. We therefore recommend small-scale validation of random samples before using a dataset for training or evaluation.}

\section{Alternative Benchmarks\label{sec:datasets}}

\changed{Whereas the Discussion synthesises the implications of our findings, this section provides guidance on selecting benchmarks and datasets for same-main-function semantic clone detection.}
\changed{No single alternative is universally preferable. Researchers should select and, where necessary, validate a benchmark against the intended clone construct, scale, language, provenance, and required positive and negative labels.}
\changed{To identify these alternatives, we conducted a literature review of relevant benchmarks and datasets.}

We will first discuss alternative benchmarks specifically designed for clone detection, and then discuss datasets that are not specifically designed for clone detection but can be used for training and evaluation. We list all the identified benchmarks and datasets and discuss each of them below.

There have been surveys that include benchmarks for clone detection.
\citet{zakerinasrabadi2023} performed a systematic literature review on source code similarity measurement and clone detection, including 12 benchmarks.
\citet{Zhang2025} surveyed machine learning approaches to code similarity measurement and discussed 11 benchmarks.
\citet{Alshabib2025} focused on cross-language clone detection, including a discussion of the BigCloneBench, OJClones, and Google Code Jam datasets.
The three surveys observe a lack of reliable and accessible datasets, which is a major limitation in the field.
Other observed problems include the lack of standardisation, i.e., papers using different benchmarks, different versions, or non-public datasets, and a lack of diversity, i.e., benchmarks usually focus on a small set of languages and are often extracted from student submissions or coding contests.

\subsection{Clone Detection Benchmarks}

There are a few benchmarks in the literature that can be used for evaluating and/or training machine learning-based code clone detection approaches, although they have been designed for the evaluation of traditional clone detection approaches.
Other benchmarks and datasets that do not include semantic clones for clone detection are not necessarily helpful in this case and are not considered.
One example is Bellon's benchmark~\citep{Bellon2007}, which, similarly to BigCloneBench, was primarily targeted at the recall of traditional clone detection tools.

\subsubsection*{OCD}
The Obfuscation, Compilation and Decompilation (OCD) dataset~\citep{Krinke2021,Ragkhitwetsagul2018} is an artificial Java code similarity benchmark that has been created using code obfuscators, compilers, and decompilers. 
The dataset is created by applying a series of transformations to the original code, including renaming variables, changing control flow, and code structure to create functionally similar but syntactically different code fragments.
\changed{The original evaluation counts 1,000 true clone pairs and 9,000 false clone pairs, created from 10 different functionalities, including self-pairs and both orientations of each pair.}

\subsubsection*{Clone Oracle}
\citet{Krutz2014} created a clone oracle dataset by having a group of multiple experts and students manually inspect a random sample of 1,536 function pairs from three open-source projects, including Apache, Python, and PostgreSQL. Then, they determined if the sampled pairs were clones of each other.
The clone oracle dataset consists of 66 clone pairs, 43 of which are Type-2 clones, 14 are Type-3 clones, and 9 are Type-4 clones.
The dataset is relatively small and not sufficient for training machine learning models, but it can be used for evaluating the performance of machine learning-based code clone detection tools along with other benchmarks.

\subsubsection*{SeSaMe}
\citet{Kamp2019} created the SeSaMe dataset of Java method pairs that are classified based on their functional similarity.
\changed{They mined JavaDoc comments from 11 Java open-source projects, then used the text similarity between the comments to select candidate method pairs for manual assessment.}
\changed{Eight judges manually assessed a selection of 900 Java method pairs for similarity of goals, operations, and effects, using disagreement, conditional agreement, and agreement ratings.}
\changed{The filtered release contains 857 assessed pairs, each accompanied by individual similarity and confidence ratings.}
\changed{We report this total without imposing a binary classification, since the numbers of positive and negative pairs depend on the chosen similarity dimensions and the treatment of conditional, uncertain, or conflicting ratings.}
Similar to the clone oracle dataset, the SeSaMe dataset is relatively small, so it may not be suitable for training machine learning-based code clone detection tools, but it can be used for their evaluation.

\subsection{Semantic Clone Detection Benchmarks}

There are three benchmarks that have been created specifically for the detection of semantic clones.

\subsubsection*{SemanticCloneBench}
The dataset~\citep{AlOmari2020} is created from answers on Stack Overflow.
The dataset contains 4,000 manually validated clone pairs in four programming languages: Java, C, C\#, and Python (with 1,000 clone pairs each).
The construction process of the benchmark includes several filtering steps, including the syntax validation using TXL~\citep{Cordy2004} to get method-level code fragments, the functionality validation based on the votes of the Stack Overflow users, the semantic clone filtering using NiCad clone detector, and the manual validation by two judges.
Due to its low number of clone pairs, the dataset is more suitable for evaluation than training machine learning models.
The benchmark has been used to evaluate the performance of code similarity tools and models in several recent studies~\citep{Arshad2022,Rabbani2022,Abid2023,Pinku2024,Awal2024,Gupta2024}.

\subsubsection*{GPTCloneBench}
The GPTCloneBench~\citep{Alam2023} dataset is created based on SemanticCloneBench and OpenAI's GPT-3 model.
It supports four programming languages similar to the SemanticCloneBench, including Java, C, C\#, and Python.
\changed{The benchmark contains 37,149 true semantic clone pairs, 19,288 pairs labelled as ``false semantic'' (Type-1/Type-2 clone pairs), and 20,770 cross-language clone pairs (Java--C\#).}
\changed{The ``false semantic'' label excludes syntactic clones from the benchmark's semantic clone category, but does not establish that the paired fragments are functionally dissimilar.}
The semantic clones are created by the GPT-3 model based on the original code fragments in  SemanticCloneBench and are filtered using NiCad to remove syntactic clones.
Then, the remaining pairs are manually validated by nine judges to ensure the functional similarity.
It has been used in recent studies to evaluate clone detection methods~\citep{Pinku2024,Zhang2024}.

\subsubsection*{FEMPD}
The FEMPD is a dataset of functionally equivalent method pairs in Java~\citep{Higo2024}.
It includes 1,342 functionally equivalent method pairs that pass the test cases of each other and are manually validated by three judges.
The dataset is used to evaluate the performance of code clone detection tools~\citep{Higo2024} and to improve clone detection accuracy~\citep{Inoue2024}.

\subsection{Coding Competitions Datasets}

There are several datasets that are created from coding competitions and online judge systems that have been used for the training and evaluation of code similarity approaches.

\subsubsection*{Google Code Jam (GCJ)}
\citet{Petrik2021} created the GCJ dataset by collecting code submissions from the Google Code Jam competition from 2008 to 2020.
It contains 332 different programming problems and 2,430,000 code submissions spanning over 20 programming languages.
\citet{Zhao2018} also collected a smaller set of the Google Code Jam dataset containing 12 programming problems for their study on deep-learning code clone detection.

\subsubsection*{CodeNet}
The CodeNet dataset~\citep{Puri2021} is a large dataset comprising code fragments in various programming languages.
It is designed for evaluating code understanding and generation tasks, including code clone detection.
It consists of code fragments from two online judge systems, AIZU and AtCoder.
\changed{AIZU's published terms, revised on 8 April 2016, permit site users to copy or modify public code for educational or research purposes and prohibit commercial use.\footnote{\url{https://judge.u-aizu.ac.jp/onlinejudge/submission_note.jsp} (accessed 23 September 2026).}}
\changed{AtCoder's terms archived on 29 April 2021 retained users' copyright and granted AtCoder a licence for advertising and notification, without expressly authorising third-party research or machine-learning reuse.\footnote{\url{https://web.archive.org/web/20210429191501/https://atcoder.jp/tos} (accessed 23 September 2026).}}
\changed{The revision of 29 June 2026 added a provision for supplying submissions and related data to third parties for machine-learning use with user consent.\footnote{\url{https://atcoder.jp/tos} (accessed 23 September 2026).}}
\changed{Neither version establishes whether separate permissions covered CodeNet's original collection.}
\changed{The CodeNet supplementary material states that IBM has the right to republish the associated third-party code under open-source licence terms~\citep[Supplementary material, Section A.2]{Puri2021}.}
\changed{We therefore recommend checking the applicable licences, permissions, and intended use before reusing the AtCoder-derived data, with institutional ethics and legal review where appropriate~\citep{Gold2022}.}
\changed{Ethics approval does not itself grant copyright permission.}

\subsubsection*{CLCDSA}
The CLCDSA dataset~\citep{Nafi2019} is a cross-language code clone dataset covering three programming languages: Java, C\#, and Python.
It is collected from AtCoder, Google Code Jam, and Coderbyte.
The dataset contains 78K solutions and has been used in a few previous studies~\citep{Nafi2019,Pinku2024}.
\changed{As with CodeNet, we do not recommend reusing the AtCoder-derived data without checking the applicable licences and permissions and obtaining institutional ethics and legal review where appropriate.}

\subsubsection*{CF-500}
\citet{Xue2022} created the CF-500 dataset from the Codeforces online judge system.
It contains 23,146 code fragments in C that cover 500 programming problems, and uses them to evaluate their semantic graph-based Type-4 clone detection approach.

\subsubsection*{Code4Bench}
\citet{Majd2019} created the Code4Bench dataset from the Codeforces online judge system, similar to the CF-500 dataset.
However, it contains a larger number of 3,421,357 code fragments in 28 programming languages such as C/C++, Java, Python, and Kotlin.
It covers 2,990 programming problems across 541 competitions.

\subsubsection*{POJ}

The POJ-104/OJClones dataset~\citep{Mou2016} is a widely used benchmark constructed from submissions to a POJ pedagogical online judge system.
The dataset contains 52,000 code samples in C/C++ that are answers to 104 programming problems.
It has been used for code classification~\citep{BenNun2018} and code similarity tasks~\citep{Ye2020a}.
Many of the papers investigated in the literature review use the POJ dataset in addition to BigCloneBench, enabled by the inclusion of both datasets in the CodeXGLUE dataset~\citep{Lu2021b}.

\subsubsection*{XLCoST}
\changed{The XLCoST benchmark~\citep{Zhu2022} is a cross-language machine learning benchmark for several code-related tasks, including code translation and code search.}
It is derived from the GeeksForGeeks website and contains code in 7 common programming languages, including C++, Java, Python, C\#, JavaScript, PHP, and C, and their natural language specifications (English).
The data is parallel across 7 languages, which makes it suitable for cross-language code clone detection.
\changed{Its program-level code-to-code data comprises approximately 123K parallel program pairs across the training, validation, and test splits.}
\changed{Our embedding analysis uses the 6,000-pair CLCCD (XLCoST) dataset derived from XLCoST by \citet{Moumoula2025}, rather than the full benchmark.}

Table~\ref{tab:alternative} presents an overview of the benchmarks 
and datasets.
It shows the names of potential semantic-clone benchmarks or datasets, their types, the languages they cover, the number of true and false clone pairs, or the number of code samples, and the type of validation performed on the clone and non-clone pairs.

\begin{table*}
	\caption{Potential Alternative Semantic Clone Benchmarks. 
	Datasets annotated with $^*$ have potential copyright issues.
	\changed{The $^\dagger$ entry counts GPTCloneBench's Type-1/Type-2 pairs labelled as ``false semantic'', not validated functionally dissimilar pairs.}}
	\label{tab:alternative}
	\resizebox{\textwidth}{!}{
	\centering
	\begin{tabular}{ll@{}r@{}rl}
	\toprule
	Clone Benchmark & Language & Clone Pairs & False clone Pairs & Validation \\
	\midrule
	OCD & Java & 1,000 & 9,000 & By construction \\
	Clone oracle & C/C++ & 66 & \changed{1,470} & Manually \\
	SeSaMe & Java & \multicolumn{2}{r}{\changed{857 assessed pairs}} & Manually \\
	SemanticCloneBench & Java, C, C\#, Python & 4,000 & - & Manually \\
	GPTCloneBench & \changed{Java, C, C\#, Python} & 37,149 & \changed{19,288$^\dagger$} & Manually \\
	FEMPD & Java & 1,342 & 852 & Manually \\
	\midrule
	Coding Competition & Language &  & Size & Validation \\
	\midrule
	Google Code Jam & 20 languages & \multicolumn{2}{r}{2.4M code samples} & Online Judge \\
	CodeNet$^*$ & 55 languages & \multicolumn{2}{r}{13.9M code samples} & Online Judge \\
	CLCDSA$^*$ & Java, C\#, Python & \multicolumn{2}{r}{78K code samples} & Online Judge \\
	CF-500 & C & \multicolumn{2}{r}{23,146 code samples} & Online Judge \\
	Code4Bench & 28 languages & \multicolumn{2}{r}{3.4M code samples} & Online Judge \\
	POJ/OJClones & C/C++ & \multicolumn{2}{r}{52,000 code samples} & Online Judge \\	
	XLCoST & 7 languages & \multicolumn{2}{r}{\changed{$\approx$123K parallel program pairs}} & Matching comments \\
	\bottomrule
	\end{tabular}
	}
\end{table*}

\subsection{Recommendation}

\changed{Because these datasets were created for different purposes and each has distinct limitations, no single recommendation applies to every use.}
\changed{We therefore organise the guidance by intended task.}

\begin{enumerate}
	\item \textit{Validating the overall performance of a clone detection approach:}
	\changed{The choice of benchmark must match the intended clone definition and evaluation metric.}
	\changed{Recall can be measured against the known true clone pairs.}
	\changed{Positive reference sets alone do not support a general precision estimate, which requires validation of reported detections or a representative evaluation set with labelled true and false pairs.}
	\changed{SemanticCloneBench does not provide false clone pairs~\citep{AlOmari2020}.}
	\changed{SeSaMe provides graded similarity judgements for goals, operations, and effects, so binary precision and recall require an explicit rule for deriving positive and negative labels~\citep{Kamp2019}.}
	\item \textit{Validating the performance of a machine learning-based code clone detection approach on a small scale:} \changed{We recommend using the GPTCloneBench and FEMPD datasets, as they have been created with a clear focus on semantic clones and have been manually validated by human judges.} \changed{GPTCloneBench's ``false semantic'' pairs require validation against the intended functional-similarity definition before they can be used as negative examples, since Type-1/Type-2 clones are not necessarily functionally dissimilar~\citep{Alam2023}.} \changed{However, the FEMPD dataset is limited to only Java, and the GPTCloneBench dataset is limited to Java, C, C\#, and Python.} The OCD dataset can also be used for this purpose, but it is an artificial dataset generated by code transformations and may not reflect real-world code similarity.
	\item \textit{Validating the performance of a machine learning-based code clone detection approach on a large scale:} We recommend using the datasets created from coding competitions, as they contain a large number of code samples that have been validated by online judges or matching comments.
	\changed{Solutions to the same programming problem can differ in their implementations, making these datasets useful candidates for evaluating functional similarity rather than only syntactic similarity~\citep{Puri2021}.}
	\changed{However, datasets created from coding competitions may not reflect the diversity of real-world code similarity.}
	\changed{Moreover, some datasets created from coding competitions may raise copyright issues, so researchers should check the applicable licences and obtain any necessary permissions from rights holders, alongside institutional ethics and legal review where appropriate.} Lastly, these coding competition datasets do not provide explicit labels for clone pairs. When using such datasets, one needs to be careful about the assumption that code fragments solving the same problem are similar, whereas those solving different problems are not. This assumption may not hold in all cases, and it is recommended to perform a critical analysis of the dataset before using it for evaluating machine learning-based code clone detection approaches.
	\item \textit{Training machine learning-based code clone detection approaches:} We recommend using the datasets created from coding competitions, as they contain a large number of code samples with validations. However, all the caveats mentioned above for using coding competition datasets to validate machine learning-based code clone detection approaches also apply to their use for training. 
\end{enumerate}

Most importantly, manually curated benchmarks for code similarity differ from datasets generated automatically, either via code transformation~\citep{Ragkhitwetsagul2018,Krinke2021} or by mining code from online judge systems~\citep{Puri2021,Mou2016,Petrik2021}.
Manually curated benchmarks suffer from subjectivity and limited size, and they target similar, but not necessarily identical, functionalities.
\changed{For accepted submissions to the same problem, online judges provide evidence of functional agreement on the tested inputs.}
\changed{Passing these tests does not establish identical functionality on all inputs.}

None of the above datasets has been scrutinised as thoroughly as BigCloneBench, and manual investigations of the datasets have been limited.
The only recommendation for the use of alternative benchmarks is that they be combined with human critical analysis to determine to what extent the assumptions about a dataset hold.

While the construction of BigCloneBench is documented in great detail, the literature analysis showed that other datasets often lack technical details about their construction.
We recommend that future datasets provide detailed documentation of their construction process to allow for critical analysis and understanding of the dataset.

\changed{Where resources permit, evaluating the same claim on multiple suitable, independently constructed and construct-aligned datasets can reveal common patterns and whether a conclusion depends on a particular dataset, although each dataset must still be critically assessed before use.}
The CloReCo platform~\citep{Burock2025} demonstrates that the community is already building more robust evaluation infrastructure.

\changed{Across these alternatives, the central lesson is that researchers must validate a benchmark's assumptions and alignment with the intended task before using it for training or evaluation.}

\section{Threats to Validity}

\changed{Our work critiques the validity of a large body of work in semantic clone detection research, and we must apply the same level of transparency and rigour to our own study.}

\subsubsection*{Conclusion Validity}

\changed{The primary threat is the reliability of our measures.}
\changed{For the stratified sample of 406 clone pairs, the initial inter-rater reliability, as measured by Gwet's AC1, showed almost perfect agreement.}
\changed{Nonetheless, judging functional similarity remains inherently subjective.}
\changed{We discussed the conflicting cases to reach a consensus, but adjudication resolves observed disagreement rather than eliminating construct uncertainty.}
\changed{The population-weighted estimate of 93.9\% has a conservative 95\% confidence interval of 91.5\%--96.2\%. This interval describes sampling uncertainty while treating the adjudicated labels as fixed and uses the maximum possible binary-outcome variance for singleton strata.}
\changed{As a sensitivity analysis independent of adjudication, we treated all 42 initial disagreements as true positives and counted only the 341 pairs independently rejected by both investigators.}
\changed{The corresponding population-weighted estimate remains 84.4\% (95\% confidence interval: 81.4\%--87.4\%).}
\changed{We also compared the human consensus labels with an LLM analysis, which showed substantial agreement, but we do not treat the LLM as an independent oracle that removes human judgement.}

\subsubsection*{Internal Validity}

The main threat is researcher bias.
\changed{As authors of a paper critical of BigCloneBench, we may be biased towards identifying problems.}
We mitigated this in the manual study by having two authors independently label the sample before any discussion.
In the literature review, we explicitly acknowledged this threat, and we employed a multi-rater classification system involving an AI, an AI-assisted human, and an unassisted human, followed by a consensus discussion.
\changed{This design provided an unassisted judgement during consensus, but it does not isolate or quantify any anchoring effect.}
We acknowledge that there may still be a margin of error in the final consensus count of 128 papers, and it should be interpreted as a well-informed estimate.

\subsubsection*{Construct Validity}

\changed{Our diagnostic experiment measured functional similarity with seven embedding models applied to source code.}
\changed{Agreement among their results reduces dependence on any single model, but does not eliminate common threats arising from shared architectures, potentially overlapping training data, or undocumented source-code exposure.}
\changed{Several models share Qwen-derived bases and may have overlapping training data, so their agreement is robustness evidence rather than independent replication.}
\changed{Direct training on BigCloneBench's labelled pairs with a similarity objective would be expected to increase agreement with its ground truth, making the observed contrast smaller rather than explaining it. Exposure to the underlying source code alone has no equally clear directional effect.}
\changed{The models' consistently higher scores on the six comparison datasets support the diagnostic interpretation of the contrast with balanced BigCloneBench, but do not establish that embedding similarity is an oracle for functional similarity.}

Functional similarity is an inherently subjective concept, leading to potential disagreements between human raters.
\changed{We operationalised it as two methods achieving the same result or performing the same task as their main overall purpose. Sharing only a subordinate functionality was insufficient.}
\changed{The protocol and consensus procedure described in \Cref{sec:manual-checking} make this interpretation explicit, but deciding whether a behavioural difference changes the task or merely its implementation remains judgement-dependent.}
\changed{The estimated mislabelling rate is conditional on this interpretation and should not be generalised to more permissive notions of functional overlap.}

In the literature review, the label of a paper's validity being ``potentially threatened'' is an interpretation of extracted information, not a direct measurement.
\changed{We mitigated this threat through independent classifications and consensus between two human reviewers, one of whom worked without LLM assistance, while recognising that consensus does not eliminate classification uncertainty.}

\subsubsection*{External Validity}

\changed{For our population estimate, we used all 406 pairs in the disproportionate stratified probability sample and weighted each functionality by its share of the eligible WT3/T4 sampling population.}
\changed{Twenty strata contain a pair added after the initial proportional allocation, while other small strata also contain only one observation. Their within-stratum variances cannot be estimated from the sample.}
\changed{We therefore use the maximum possible binary-outcome variance for every singleton stratum, making the reported interval conservative with respect to this limitation.}
\changed{We did not sample the \texttt{sample-sample} and \texttt{sample-tagged} WT3/T4 categories, so the estimated 93.9\% mislabelling rate and its confidence interval apply only to the sampled \texttt{tagged-tagged} category.}

\subsubsection*{Threats Introduced by the Use of an LLM}
\changed{The LLM had two distinct roles: a post-hoc challenge to the clone-pair labels and extraction assistance in the literature review.}
\changed{For the clone-pair comparison, results may be sensitive to the prompt, the fixed order of the snippets, and the exploratory selection of the model and prompt. We archive the final prompt and runs, but did not retain the exploratory comparison or conduct a systematic order-sensitivity analysis.}
\changed{Four identical runs and a one-pair variation in the fifth establish only within-configuration repeatability, not correctness, while the dated proprietary API may not remain available or reproduce the exact outputs.}
\changed{The surviving configuration record also lacks the seed setting, output-token limit, retry policy, access dates, and original request script. These omissions are stated in the artefact rather than supplied retrospectively.}
\changed{The model received neither BigCloneBench labels nor functionality tags, but exposure to IJaDataset source code or related benchmark material remains possible: \citet{zhou2025} report 55.7\% data leakage for BigCloneBench source code, and neither agreement nor disagreement with our labels can rule out memorisation.}
\changed{Because the authors saw the explanations before reconsidering the 12 disagreements, automation or confirmation bias may have affected the two accepted changes. We therefore retain the pre-LLM human estimate as the primary result and report the post-LLM estimate only as a sensitivity analysis.}
\changed{In the literature review, long PDFs may cause the ``lost in the middle'' effect \citep{liu2023}, while model suggestions may anchor the assisted reviewer. Requiring consensus with a second reviewer who worked without those suggestions reduces, but does not eliminate, these threats.}

\subsubsection*{Ethical and Legal Considerations}
\changed{The discussion of AtCoder-derived data identifies the need to check permissions for the intended use, rather than establishing that the historical collection or subsequent reuse was unlawful.}
A comprehensive ethical and legal audit of all the mentioned alternative benchmarks was not conducted and is outside the scope of our study, which focuses on the investigation of BigCloneBench.
\changed{For the underlying IJaDataset code, our artefact records the sampled method identifiers and source snippets, but these provenance records do not establish comprehensive licence compliance or ethical approval.}
We acknowledge that ethical and legal considerations are crucial in the creation and usage of datasets, and we encourage future work to address these issues more comprehensively.

\section{Conclusions}

\changed{Under our operational definition of functional similarity, the population-weighted analysis of our stratified probability sample of 406 WT3/T4 clone pairs estimated that 93.9\% of the eligible pairs in the sampled category were mislabelled as clones (conservative 95\% confidence interval: 91.5\%--96.2\%).}
\changed{A sensitivity analysis that does not depend on resolving disagreements between the investigators still estimated 84.4\% (95\% confidence interval: 81.4\%--87.4\%).}
\changed{Although the precise magnitude depends on the interpretation of functional similarity, both analyses indicate that the WT3/T4 ground truth is unsuitable for evaluating whether methods have the same main functionality.}
\changed{This estimate applies only to \texttt{tagged-tagged} WT3/T4 pairs: this study does not assess the exemplar-containing pair categories or uses governed solely by textual clone-category definitions.}
Creating a dataset on the scale of BigCloneBench is challenging.
\changed{The dataset's contributions to traditional clone detection are significant and have enabled research in this area for more than a decade.}

The discussed issues are a larger threat to machine learning approaches, where the ground truth is used to learn whether a pair of methods are clones or not.
\changed{Because machine learning approaches commonly use WT3/T4 clone pairs, conclusions drawn solely from these evaluations may be threatened by the ground-truth misalignment.}
\changed{In our investigation of 179 papers that use BigCloneBench as a dataset, our final consensus identified 128 papers that used BigCloneBench to evaluate semantic clone detection and whose results may therefore be affected by the mislabelling.}
\changed{The literature-classification process also showed that LLM suggestions can create a risk of anchoring, although the design does not isolate that effect from differences between reviewers.}
\changed{Independent classification by an unassisted human reviewer was therefore important to the final consensus of 128 papers (71.5\%).}
\changed{The LLM's highly skewed classifications also produced low Cohen's Kappa despite moderate observed agreement.}

\changed{As such, these 128 papers often report high F1 scores (e.g., above 0.9), but such scores on BigCloneBench do not establish whether the approaches learned functional similarity or relied on dataset-specific shortcuts.}
Moreover, we observed that some machine learning approaches address a large number of unlabelled code pairs by changing or replacing the ground truth, making incorrect assumptions about methods for different functionalities. 
By doing so, the approaches generate large numbers of code pairs that are incorrectly labelled as not being clones.

\changed{The widespread misuse of BigCloneBench's WT3/T4 subset as ground truth for same-main-function similarity, without validating its labels for that task, has harmed semantic clone detection by weakening its evidence base: a substantial body of reported results now requires re-evaluation.}
\changed{We hope to raise awareness of the flawed WT3/T4 ground truth and how its construction affects the validity of evaluations.}
We encourage a more critical and responsible use of benchmarks and datasets and ask for more rigorous validation during dataset creation and use in the future.

\changed{We therefore recommend against using BigCloneBench's WT3/T4 subset, or datasets derived from it without independent label validation, as ground truth for machine learning approaches intended to detect same-main-function similarity.}

\section{Declarations}

\subsection{Funding}
Chaiyong Ragkhitwetsagul was financially supported by the Office of the Permanent Secretary, Ministry of Higher Education, Science, Research and Innovation (RGNS64-143).

\subsection{Ethical approval}
Not applicable

\subsection{Informed consent}
Not applicable

\subsection{Author Contributions}
\begin{itemize}
    \item \textbf{Jens Krinke:} Conceptualization, Data curation, Formal analysis, Investigation, Methodology, Project administration, Resources, Software, Supervision, Validation, Visualization, Writing - original draft, Writing - review \& editing
	\item \textbf{Chaiyong Ragkhitwetsagul:} Conceptualization, Data curation, Formal analysis, Funding acquisition, Investigation, Methodology, Project administration, Resources, Validation, Visualization, Writing - original draft, Writing - review \& editing
\end{itemize}

\subsection{Data Availability Statement}
We provide the following data as supplementary material on Zenodo~\citep{Krinke2025Zenodo} and GitHub~\citep{BigCloneBenchSampleValidation}:
\begin{itemize}
\item \changed{The aggregate results of the exploratory embedding analysis, including AUC, optimal thresholds, and MCC and F1 at the optimal thresholds and at 0.5.} \changed{Pair-level scores and results using BigCloneBench-derived thresholds are not included.}
\item The subset of 406 clone pairs used for the manual investigation.
\item The protocol for the manual investigation.
\item The results of the manual investigation.
\item \changed{The LLM-based post-hoc comparison artefact, including the pair-level majority table and offline reproducer.}
\item The code used for the LLM-based analysis of the literature review.
\item \changed{The literature-review bibliography and aggregate classification and agreement results.}
\end{itemize}
\changed{Individual evaluative classifications of papers and identifiable literature LLM responses are retained internally because they could be interpreted as broader judgements about individual papers than the analysis supports.}
\changed{The individual literature classifications and LLM responses are available from the authors on request.}
\changed{Consequently, the literature agreement statistics cannot be independently recalculated from the public supplementary files alone.}
The protocol for the manual investigation and the prompts used for the literature search are provided in the appendix for the purpose of the paper review.

\subsection{Conflict of Interest}
Not applicable

\subsection{Clinical trial number}
Not applicable

\section*{Acknowledgements}

We would like to thank the authors of BigCloneBench for their feedback and discussion on a draft of this paper.
We asked the authors of the papers we explicitly mention~\citep{Yu2019,Wang2020,Ji2021,Zhang2023,Chunyan2023,Yu2023,Lu2021b} to comment on our findings, but we did not receive any feedback. 
This work was financially supported by the Office of the Permanent Secretary, Ministry of Higher Education, Science, Research and Innovation, Thailand.

\bibliographystyle{spbasic}

\begin{onecolumn}
\appendix
\ifdefined\isarxiv
\else
\emph{The appendices are mainly for the purpose of the paper review and are part of the accompanying artefact.
They are only provided as supplementary material for the reviewers to ease the review.}
\fi

\section{Protocol for the Manual Investigation}

The following is the protocol for the manual investigation of the methods.
\changed{Both authors followed the protocol during the manual assessment of the 406 sampled WT3/T4 true clone pairs.}

\subsection{Instructions}

Consider the two snippets (methods) and answer the two questions:

\begin{enumerate}
\item Does each of the two snippets implement the functionality as its main or only purpose?
\item Are the two snippets (methods) Type-3 or Type-4 clones?
\end{enumerate}

\subsection{Question 1}

The functionality of each method should be checked against the specification as found in the respective table.
The two methods should be labelled as follows:
\begin{enumerate}
\item \textbf{major} The method implements the whole functionality as its main or only purpose.
	It implements the functionality true to the specification.
\item \textbf{incomplete} The method implements only part of the functionality as its main or only purpose.
	Some part of the functionality is missing and likely implemented in another method that calls the method of interest, or in another method that the method of interest calls.
\item \textbf{minor} The method uses the functionality, but has a different main purpose.
\item \textbf{wrong} The method does not implement the functionality true to the specification.
\item \textbf{test} The method implements the functionality as part of a test.
\end{enumerate}

\subsection{Question 2}

We rely on the original definition of Type-3 and Type-4 clones:

A Type-3 clone is
\begin{itemize}
\item \textit{A clone with very similar source code, but with small changes made to the code to tailor it to some new function}~\citep{Carter1993}.
\item \textit{A copied fragment with further modifications such as changed, added, or removed statements, in addition to variations in identifiers, literals, types, whitespace, layout, and comments}~\citep{Roy2009}.
\end{itemize}

A Type-4 clone is
\begin{itemize}
\item \textit{A functionally identical clone, possibly with the originator unaware that there is a function already available that accomplishes essentially the same function}~\citep{Carter1993}.
\item \textit{Two or more code fragments that perform the same computation but are implemented by different syntactic variants}~\citep{Roy2009}.
\end{itemize}
However, Type-4 clones, as defined above, require identical functionality or the same computation.
Such a definition is not decidable in general, and we only require that the two fragments are functionally similar.
As a definition for similar functionality, we use the following:
\begin{quote}\it
	Two code fragments are considered functionally similar when they achieve the same result or perform the same task, even if they differ in syntax, structure, or the specific steps they take to accomplish that task.
\end{quote}

In the definition from Roy et al.~\citep{Roy2009}, we drop the requirement that they are implemented by different syntactic variants (due to the automatic classification, there should be no clones with similar syntactic variants in the sample).
There should be no Type-3 pair, as all Type-3 pairs should have been filtered out by the automatic filter based on dissimilarity.
If there is, the pair needs close attention.

Based on the labels of the two methods, usually, the second question is only true for a major/major pair.
However, there may be cases where the two methods are labelled differently, but are still clones of each other.
Such cases need close attention.

\section{Instructions for the LLM during the Literature Search}

\subsection{First Prompt}

{\footnotesize
\begin{verbatim}
**Role:**  
You are an experienced Software Engineering Researcher specializing in Clone
Detection, with expertise in empirical studies and dataset evaluation.  

**Context:**  
You are analyzing a research paper that uses **BigCloneBench**, a large-scale
dataset for evaluating clone detection tools. Your goal is to understand how
the dataset is used in the paper, verify key details, and critically assess its
validity.  

BigCloneBench exists in two versions:  
- The **old version**: ~6.1 million true clone pairs, ~258,000 false clone
  pairs, ~60,000 code snippets, **10 functionalities**.  
- The **new version**: ~8.9 million true clone pairs, ~288,000 false clone
  pairs, ~74,000 code snippets, **43 functionalities**.
Note that BigCloneBench is often shortened to BCB in research papers.

You have access to the research paper and need to **extract relevant details,
answer specific questions, and analyze the validity of the study's findings**.  

---

### **Tasks:**  

1. **Extract Key Metadata:**  
   - **Title, authors, and publication year** of the uploaded paper. 
     The paper number is PAPER_NUMBER.  

2. **Summarize the Paper:**  
   - Provide a **concise one-paragraph summary** covering **objectives, 
     methodology, key findings, and conclusions**.

3. **Extract Dataset Usage:**  
   - Provide a **concise summary** of which datasets are used for evaluation.

4. **Analyze BigCloneBench Usage:**  
   Answer each of the following research questions with a **justified
   explanation** and a **direct quote** with a corresponding page number.

   - **A:** Is the paper a **systematic literature review (SLR) or a survey**?  
   - **B:** Does the paper present a **novel clone detection or clone search 
     approach** or **evaluate existing approaches**?  
   - **C:** Does the paper **use BigCloneBench for evaluation**?  
   - **D:** Is **BigCloneBench used as ground truth for training a machine
     learning approach**?  
   - **E:** Which **version of BigCloneBench (old/new)** has been used?  
   - **F:** Has the **ground truth of BigCloneBench been filtered/modified**?
     If so, what is the **size of the subset used**?  
   - **G:** Has the **WT3/T4 subset been excluded** from evaluation?
     If filtered, was **WT3/T4 part of the subset**?  
   - **H:** Ignoring filtering, has the ground truth otherwise **been changed,
     extended, or enriched**?  
   - **I:** Has the paper used a **subset created by previous work**?  
   - **J:** Has the paper **validated or manually investigated**
     BigCloneBench's ground truth?  
   - **K:** Does the paper **cite "BigCloneBench Considered Harmful for Machine
     Learning"?**  
 
   Each answer should be in the following format:
   - **Q <X>:** Question text?  
   - **A:** Answer (Yes/No).
   - **Explanation:** Justified explanation.
   - **Quote:** Relevant quote with page number.

5. **Critical Analysis -- Impact of New Findings:**  
   - **Recent research** found that **93.35% of WT3/T4 pairs in BigCloneBench
   	 are not clones**.  
   - **Assess the impact of this finding** on the validity of the paper’s
     results.  
     - **L:** Does this finding **weaken or invalidate** any claims in the
       paper?  
     - How does this affect **methodology, conclusions, or generalizability**?
\end{verbatim}
}

\subsection{Second Prompt}
{\footnotesize
\begin{verbatim}
### **Final Summary: Tabular Assessment**  

After answering all the questions, summarize the findings in the following CSV
table format:  

Question, A, B, C, D, E, F, G, H, I, J, K, L  
Paper PAPER_NUMBER,    ,    ,    ,    ,    ,    ,    ,    ,    ,    ,    ,      

**Instructions:**  
- **Use only "Yes" or "No"** whenever possible. 
  Other words that can be used are "Old", "New", "Not specified",
  "Potentially", or "N/A" as applicable.
- If additional clarification is needed, provide a **brief note below the
  table**.  
- Ensure that all responses are **consistent with the extracted data from the
  paper**.
\end{verbatim}
}

\end{onecolumn}

\end{document}